\documentclass[twocolumn,aps,prb,amsmath,floatfix]{revtex4}
\usepackage{graphicx}
\begin{document}
\title{Multisite versus multiorbital Coulomb correlations studied  
within finite-temperature exact diagonalization dynamical mean-field theory}
\author{A.~Liebsch$^1$, H. Ishida$^2$, and J. Merino$^3$} 
\affiliation{$^1$Institut f\"ur Festk\"orperforschung, 
             Forschungszentrum J\"ulich, 
             52425 J\"ulich, Germany \\
             $^2$College of Humanities and Sciences, Nihon University,~Tokyo 156, 
             Japan\\
             $^3$Departamento de F\'\i sica Te\'orica de la Materia Condensada,
             Universidad Aut\'onoma de Madrid, Madrid 28049, Spain}
 \begin{abstract}
The influence of short-range Coulomb correlations on 
the Mott transition in the single-band Hubbard model at half-filling 
is studied within cellular dynamical mean field theory for square and
triangular lattices.  Finite-temperature exact diagonalization is used to
investigate correlations within two-, three-, and four-site clusters.      
Transforming the non-local self-energy from a site basis to a molecular
orbital basis, we focus on the inter-orbital charge transfer between
these cluster molecular orbitals in the vicinity of the Mott transition. 
In all cases studied, the charge transfer is found to be small, 
indicating weak Coulomb induced orbital polarization despite sizable 
level splitting between orbitals. These results demonstrate that 
all cluster molecular orbitals take part in the Mott transition and that 
the insulating gap opens simultaneously across the entire Fermi surface.
Thus, at half-filling we do not find orbital-selective Mott transitions,
nor a combination of band filling and Mott transition in different orbitals. 
Nevertheless, the approach towards the transition differs greatly between
cluster orbitals, giving rise to a pronounced momentum variation along the
Fermi surface, in agreement with previous works.
The near absence of Coulomb induced orbital polarization in 
these clusters differs qualitatively from single-site multi-orbital studies 
of several transition metal oxides, where the Mott phase exhibits nearly 
complete orbital polarization as a result of a correlation driven 
enhancement of the crystal field splitting.
The strong single-particle coupling among cluster orbitals in the 
single-band case is identified as the source of this difference. 
\\
\mbox{\hskip1cm}  
	electron systems 
\end{abstract}
\maketitle

\section{Introduction}

Considerable progress has recently been achieved in the understanding of the
Mott transition in a variety of transition metal oxides.\cite{kotliar06}
Whereas density functional theory in the local density approximation (LDA) 
predicts many of these materials to be metallic, the explicit treatment of 
local Coulomb interactions via dynamical mean field theory (DMFT)\cite{dmft}
correctly yields insulating behavior for realistic values of the on-site 
Coulomb energy $U$. In the metallic phase, the non-cubic structure of some 
of these systems gives rise to non-equivalent, partially filled subbands 
that are split by a crystal field and exhibit orbital dependent electron 
occupancies. The hallmark of the Mott transition of these oxides is that 
orbital polarization can be greatly increased by Coulomb correlations 
and that the insulating phase is nearly completely orbitally polarized. 
For instance, in the case of LaTiO$_3$, the $e'_g$ bands are pushed above 
the Fermi level and the remaining singly occupied $a_g$ subband 
is split into lower and upper Hubbard bands.\cite{pavarini,prb08} 
In the case of V$_2$O$_3$, Coulomb correlations push the 
$a_g$ band above the Fermi level, and the doubly degenerate 
$e'_g$ subbands exhibit a Mott gap.\cite{keller,poteryaev} 
Also, in the insulating phase of Ca$_2$RuO$_4$, 
the $d_{xy}$ like band is completely filled and the $d_{xz,yz}$ like 
subbands are split into Hubbard bands.\cite{anisimov,prl07} 
The common feature of the Mott transition in these materials is that
the effective band degeneracy is reduced from three to two or one, 
so that the critical Coulomb energy is lower than it would be if the 
$t_{2g}$ bands were fully degenerate. 
On the other hand, other materials can exhibit a quite different behavior. 
For instance, orbital polarization in BaVS$_3$ was shown to decrease with 
increasing local Coulomb interaction.\cite{lechermann}
Also, the Mott transition in LaVO$_3$ and YVO$_3$ occurs before orbital 
polarization is complete.\cite{ray} 
Moreover, in a hypothetical tetragonal structure of LaTiO$_3$,
relevant for heterostructures, the Mott phase is reached when 
$n_{xz,yz}$ approaches $1/4$ and $n_{xy}$ vanishes.\cite{tetra}  
Finally, the possibility of so-called orbital selective Mott transitions
in multi-band systems has been discussed extensively in the literature.
\cite{anisimov,koga,prb04,song,inaba,costi}
These different trends underline the remarkably rich physics of
Mott transitions in multi-orbital materials.

The aim of this work is to investigate the relationship between Coulomb 
correlations in single-site multi-orbital systems as described above to
those occurring within a single band when inter-site Coulomb correlations 
are taken into account. The influence of short-range correlations on
the nature of the Mott transition is currently of great interest and 
has been studied by many groups.
\cite{tohyama,preuss,moreo,hettler,senechal,lich,huscroft,moukouri,%
kotliar01,imai,maier2,onoda,kyung,potthoff,senechal1,parcollet,capone,%
senechal2,civelli,maier,kyung2,capone2,kyung3,merino,zhang,ohashi,saheb,%
park,koch,ferrero,gull,lee,balzer,senechal3,Ferrero}
Here we examine the role of correlation
driven orbital polarization in the vicinity of the Mott transition.   
For example, it is well known that in a minimal two-site cluster model,
\cite{ferrero} which permits explicit treatment of short-range Coulomb 
correlations in an isotropic square lattice, the Green's function 
and self-energy become diagonal if one transforms the site basis to a 
diagonal bonding - antibonding molecular orbital basis. 
In a four-site cluster model, diagonality is obtained by 
transforming sites to cluster molecular orbitals characterized by 
$\Gamma=(0,0)$, $X=(\pi,0),(0,\pi)$, and $M=(\pi,\pi)$.\cite{park} 
The molecular orbital components of the self-energy provide 
qualitative information on the importance of correlations in 
the corresponding sections of the Brillouin Zone. 
In the case of an isotropic triangular lattice, Green's function and 
self-energy can be diagonalized by an analogous transformation to 
molecular orbitals appropriate for a three-site cluster.\cite{koch}        
The question then arises whether these cluster molecular orbitals in 
the single-band case obey a similar scenario as the multi-orbital  
systems mentioned above.

Since an approximate momentum variation of the lattice self-energy in 
these models can be derived from a linear superposition of the respective 
molecular orbital components of the cluster self-energy, the effect of
correlation enhanced orbital polarization is of direct relevance for the
question of whether the Mott gap opens uniformly across the Fermi
surface, or whether it opens first in certain regions of the Brillouin 
Zone (e.g. near the so-called hot spots) and only at larger $U$ in the 
remaining regions (the so-called cold spots). The latter picture would
be analogous to the orbital selective Mott transition which can occur
within single-site DMFT treatments of certain multi-band systems.
\cite{anisimov,koga,prb04,song,inaba,costi} Another possibility, 
analogous to multi-orbital materials such as LaTiO$_3$, V$_2$O$_3$, 
and Ca$_2$RuO$_4$, is that a subset of cluster orbitals could exhibit 
a genuine Mott transition, while the remaining ones are pushed above 
or below the Fermi level at about the same critical $U$.

To account for inter-site correlations we use DMFT combined with finite 
temperature exact diagonalization (ED).\cite{ed} 
It was recently shown\cite{perroni} that this method can be generalized 
to multi-band materials by computing only those excited states of the 
impurity Hamiltonian that are within a narrow range above the ground state, 
where the Boltzmann factor provides the convergence criterion. Exploiting 
the sparseness of the Hamiltonian, these states can be computed very 
efficiently by using the Arnoldi algorithm.\cite{arnoldi} Higher excited 
states enter via Green's functions which are evaluated using the Lanczos 
method. This approach has proved to be highly useful for the study of 
strong correlations in several transition metal oxides.
\cite{prb08,perroni,naco,tetra,v2o3-al} An important feature of ED/DMFT is 
that low temperatures and large Coulomb energies can be reached.        
The adaptation of single-site multi-orbital ED to multi-site single-band
systems is discussed in detail below. In particular, we introduce a 
mixed site - molecular orbital basis which permits a more flexible 
and more accurate projection of the lattice Green's function onto the 
cluster than in a pure site representation.  Previous multi-site ED/DMFT
studies focussed on $T=0$.\cite{kyung,civelli,kyung2,capone2,kyung3,merino,zhang}
 The extension to finite $T$ discussed here
is especially useful for the evaluation of the $T/U$ phase diagram.

The main result of this work is that in all cluster models studied 
here for half-filled square and triangular lattices, there is little
enhancement of orbital polarization in the vicinity of the Mott 
transition. Thus, despite sizable level splitting 
between these cluster orbitals, they all exhibit Mott gaps at the
same critical Coulomb energy. As a consequence, the Mott gap in these
models opens uniformly across the Fermi surface. For the square lattice
we show explicitly that the Mott gap at the cold spot $M/2=(\pi/2,\pi/2)$
of the Brillouin Zone is driven by Coulomb correlations at the hot spot 
$X=(\pi,0)$. Therefore, there is
no orbital selective Mott transition. Moreover, there is no evidence
for the combination of partial band filling and Mott transition in
remaining subbands that is characteristic of single-site DMFT 
treatments of the multi-orbital materials LaTiO$_3$, V$_2$O$_3$, 
and Ca$_2$RuO$_4$, as mentioned above.
         
The outline of this paper is as follows. Section II discusses the 
theoretical aspects of our cluster ED/DMFT implementation of finite 
temperature exact diagonalization. Section III provides the results 
for the two-site and four-site clusters of the square lattice, and 
the three-site cluster of the isotropic triangular lattice.   
In Section IV we briefly discuss analogies and differences between 
these multi-site correlation effects and those investigated previously 
in single-site DMFT treatments of multi-orbital materials.   
The conclusions are presented in Section V.         
            
\section{Multi-Site ED/DMFT} 

Let us consider the single-band Hubbard model 
\begin{equation}
   H =  -t \sum_{\langle ij\rangle\sigma} ( c^+_{i\sigma} c_{j\sigma} + H.c.) 
                     + U \sum_i n_{i\uparrow} n_{i\downarrow} 
\end{equation}
where the sum in the first term extends over nearest neighbor sites.
The hopping integral $t$ will be set equal to unity throughout this paper.
Thus, the band widths of the square and triangular lattices are $W=8$ and 
$W=9$, respectively. 
Within cellular dynamical mean field theory (CDMFT)\cite{kotliar01,maier} 
the interacting lattice Green's function in the cluster site basis is given  by
        \begin{equation}
            G_{ij}(i\omega_n) = \sum_{\vec k} \left( i\omega_n + \mu - t(\vec k) - 
                   \Sigma(i\omega_n)\right)^{-1}_{ij} \label{G}
          \end{equation}
where the $\vec k$ sum extends over the reduced Brillouin Zone, 
$\omega_n=(2n+1)\pi k_BT$ are Matsubara frequencies and $\mu$ is the chemical
potential. The lattice constant is unity.
$t(\vec k)$ denotes the hopping matrix for the superlattice  and 
$\Sigma(i\omega_n)$ represents the cluster self-energy matrix.
To make contact to other recent works,\cite{parcollet,zhang,park,gull,ferrero} 
we consider here the paramagnetic metal insulator transition.

In the site basis, the Green's functions for two-site, three-site and 
four-site clusters have the structure
        \begin{eqnarray}
           G^{(2)} &=& \left( \begin{array}{ll}
                               a & b  \\
                               b & a \\
                                    \end{array} \right)\label{G2}\\ 
           G^{(3)} &=& \left( \begin{array}{lll}
                               a & b & b \\
                               b & a & b \\
                               b & b & a \\  
                                  \end{array} \right)\label{G3}  \\ 
           G^{(4)} &=& \left( \begin{array}{llll}
                               a & b & b & c \\
                               b & a & c & b \\
                               b & c & a & b \\
                               c & b & b & a \\
                                  \end{array} \right)\label{G4}
         \end{eqnarray}
with $a=G_{11}$, $b=G_{12}$ and $c=G_{14}$. Site labels in the square lattice
refer to $1\equiv(0,0),\ 2\equiv(1,0),\ 3\equiv(0,1),\ 4\equiv(1,1)$ and in the 
triangular lattice to $1\equiv(0,0),\ 2\equiv(1,0),\ 3\equiv(1/2,\sqrt3/2)$.
The superscript denotes the cluster size $n_c$ in the square lattice 
        ($n_c=2$ or $n_c=4$) or triangular lattice ($n_c=3$), respectively.   
        In the site bases, the corresponding self-energy matrices 
        $\Sigma^{(n_c)}(i\omega_n)$ have the same symmetry properties
        as the Green's functions. 

A key aspect of DMFT is that, to avoid double-counting of Coulomb 
interactions in the quantum impurity calculation, it is necessary to 
remove the self-energy from the cluster in which correlations are 
treated exactly. This removal yields the Green's function 
        \begin{equation}
         G_0(i\omega_n) = [G(i\omega_n)^{-1} + \Sigma(i\omega_n)]^{-1} . 
           \label{G0}
        \end{equation}
These matrices also exhibit the symmetry properties specified above.

For the purpose of the ED calculations it is convenient to transform the 
site bases into molecular orbital bases in which the Green's functions and
self-energies become diagonal. For the two-site cluster, molecular orbitals 
are given by the bonding - anti-bonding combinations 
        $\phi_{1,2}=(|1\rangle\pm|2\rangle)/\sqrt2$.
        For the four-site cluster, they are formed by the plaquettes
        $\phi_1=(|1\rangle+|2\rangle+|3\rangle+|4\rangle)/2$, 
        $\phi_2=(|1\rangle+|2\rangle-|3\rangle-|4\rangle)/2$,         
        $\phi_3=(|1\rangle-|2\rangle+|3\rangle-|4\rangle)/2$, 
        $\phi_4=(|1\rangle-|2\rangle-|3\rangle+|4\rangle)/2$.
        Finally, for the three-site cluster of the triangular lattice they
        can be written as: 
        $\phi_1=(  |1\rangle+|2\rangle+|3\rangle)/\sqrt3$, 
        $\phi_2=(-2|1\rangle+|2\rangle+|3\rangle)/\sqrt6$, 
        $\phi_3=(            |2\rangle-|3\rangle)/\sqrt2$.  
 In theses cluster molecular orbital bases, the above 
 Green's functions take the form 
         \begin{eqnarray}
         G^{(2)} &=&  \left( \begin{array}{cc} 
                               a+b & 0  \\
                               0   & a-b \\
                                    \end{array} \right)\label{g2}\\
         G^{(3)} &=& \left( \begin{array}{ccc}
                               a+2b & 0   & 0 \\
                               0    & a-b & 0 \\
                               0    & 0   & a-b \\
                                  \end{array} \right) \label{g3} \\
         G^{(4)} &=& \left( \begin{array}{cccc}
                               a+2b+c & 0   & 0   & 0 \\
                               0      & a-c & 0   & 0 \\
                               0      & 0   & a-c & b \\
                               0      & 0   & 0   & a-2b+c \label{g4} \\
                                  \end{array} \right)
         \end{eqnarray}
         The self-energies $\Sigma(i\omega_n)$ and Green's functions 
         $G_0(i\omega_n)$ can be diagonalized in the same fashion. 
         We denote these elements as $G_m(i\omega_n)$, $\Sigma_m(i\omega_n)$
         and $G_{0,m}(i\omega_n)$.

\begin{figure} 
\begin{center}
\includegraphics[width=4.5cm,height=6.5cm,angle=-90]{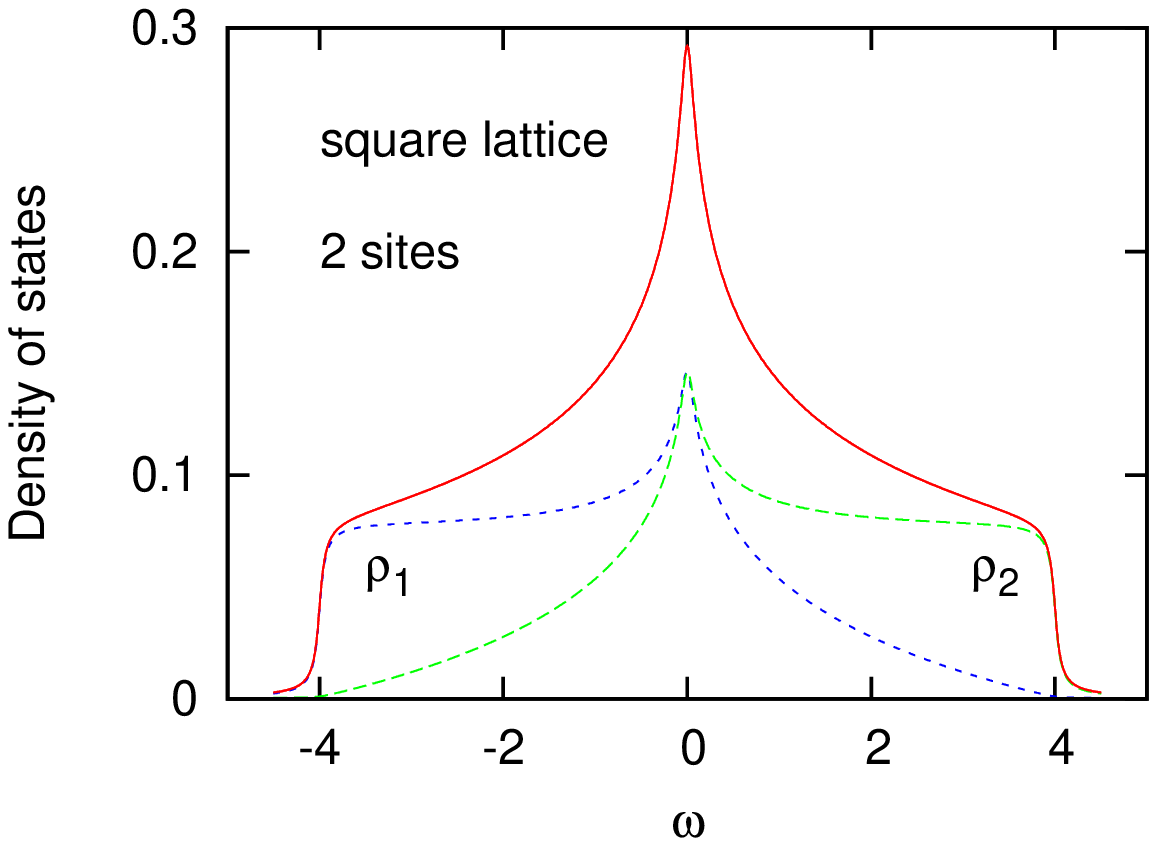}
\includegraphics[width=4.5cm,height=6.5cm,angle=-90]{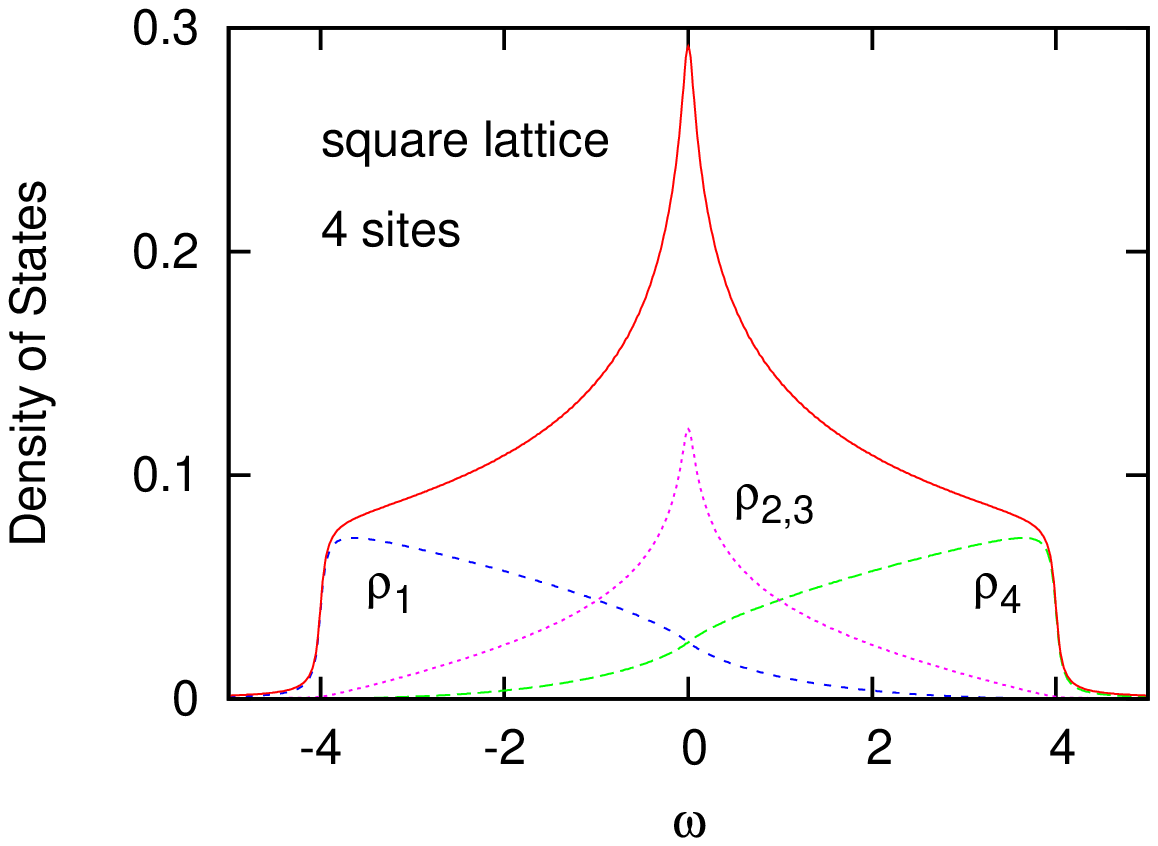}
\includegraphics[width=4.5cm,height=6.5cm,angle=-90]{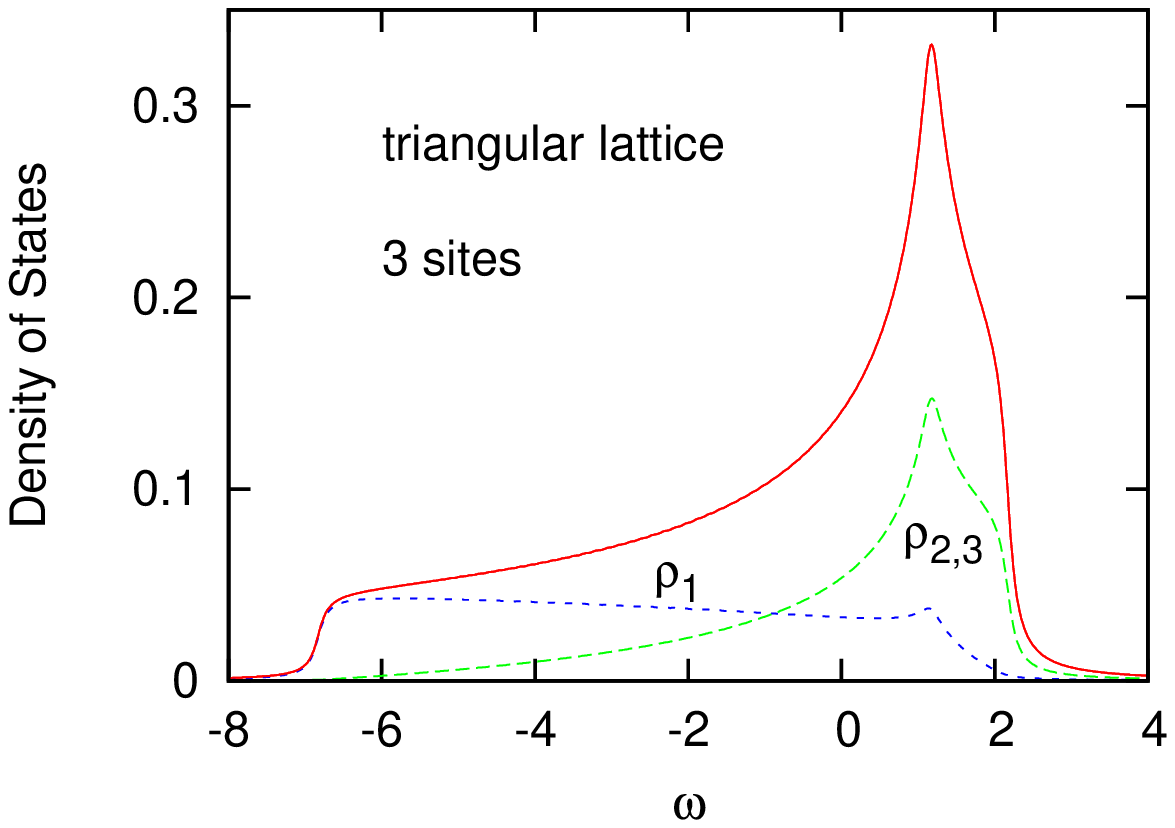}
\end{center}
\caption{(Color online)
Total density of states $\rho(\omega)$ and molecular orbital components 
$\rho_m(\omega)$ for two-site and four-site clusters of square lattice 
(top panels), and of three-site cluster of triangular lattice (bottom panel). 
For clarity, the molecular orbital components are divided by $n_c$.  
}\end{figure}

         In the site basis, the local density of states in the non-interacting 
         limit is given by
         \begin{eqnarray}
              \rho_{ii}(\omega) &=& -\frac{1}{\pi}\,{\rm Im}\, G_{ii}(\omega)
         \end{eqnarray}  
         with $\Sigma=0$. Since we consider isotropic clusters, all sites
         are equivalent, so that $\rho_{ii}(\omega)$ coincides with the density
         of states $\rho(\omega)$. In the molecular orbital basis, the density 
         of states components $\rho_m(\omega)$ have different shapes and different
         centroids, analogous to the crystal field split density of states 
         components of many transition metal oxides.    
         Figure~1 shows these densities for the two-site and four-site clusters
         of the square lattice and the three-site cluster of the triangular
         lattice, as described above.  According to Eqs.~(\ref{g2}-\ref{g4}), 
         the cluster molecular orbital densities of states are given by
         \begin{eqnarray}
         \rho^{(2)}_{1  } &=& \rho^{(2)}_{11} + \rho^{(2)}_{12} \label{rho2} \\
         \rho^{(2)}_{2  } &=& \rho^{(2)}_{11} - \rho^{(2)}_{12} \nonumber \\
         \rho^{(3)}_{1  } &=& \rho^{(3)}_{11}+2  \rho^{(3)}_{12} \label{rho3} \\
         \rho^{(3)}_{2} = \rho^{(3)}_{3} &=& \rho^{(3)}_{11} -  \rho^{(3)}_{12} \nonumber \\
         \rho^{(4)}_{1} &=& \rho^{(4)}_{11} +2\rho^{(4)}_{12}+\rho^{(4)}_{14}\label{rho4}\\
         \rho^{(4)}_{2} =\rho^{(4)}_{3} &=& \rho^{(4)}_{11}  -\rho^{(4)}_{14} \nonumber \\
         \rho^{(4)}_{4} &=& \rho^{(4)}_{11}- 2\rho^{(4)}_{12}+\rho^{(4)}_{14}\nonumber  
         \end{eqnarray}
         where $\rho^{(n_c)}_{ij}$ are the site components for cluster $n_c$.

         From these cluster molecular orbital densities of states an approximate
         momentum variation across the Brillouin Zone can be constructed (see below).
         For instance, in the case of the square lattice with $n_c=4$, densities 
         associated with the high-symmetry points of the original lattice are given by 
         $\rho_\Gamma(\omega)=\rho_1(\omega)$, 
         $\rho_X(\omega)=\rho_{2}(\omega)=\rho_{3}(\omega)$  and 
         $\rho_M(\omega)=\rho_4(\omega)$. 
         At $M/2=(\pi/2,\pi/2)$, the density of states corresponds to the local
         density $\rho(\omega)=\rho_{11}(\omega)=(\rho_\Gamma+\rho_M+2\rho_X)/4$. 
         Note, however, that all molecular 
         orbital densities extend across the entire band width. Thus, they are not 
         identical with those sections of the local density of states that originate
         in momentum regions surrounding the high-symmetry points, as would be the 
         case in the dynamical cluster approximation (DCA).\cite{hettler,maier}

             
We now project the Green's function $G_0(i\omega_n)$ defined in Eq.~(\ref{G0}) 
onto a cluster consisting of $n_c$ impurity levels and $n_b$ bath levels.
The total number of levels is $n_s=n_c+n_b$.
In the site basis we have
\begin{eqnarray}
 G_0(i\omega_n) &\approx&  G^{cl}_0(i\omega_n) \nonumber\\
                &=& \left( i\omega_n + \mu - h -  \Gamma(i\omega_n) \right)^{-1} 
\end{eqnarray}
where $h$ is the non-interacting impurity cluster Hamiltonian and $\Gamma(i\omega_n)$
the hybridization matrix describing the coupling between impurity cluster and bath.
Thus,  
        \begin{eqnarray}
           h^{(2)} &=& \left( \begin{array}{ll}
                               \epsilon_0 & t  \\
                               t & \epsilon_0  \\
                                    \end{array} \right)\\ 
           h^{(3)} &=& \left( \begin{array}{lll}
                               \epsilon_0 & t & t \\
                               t & \epsilon_0 & t \\
                               t & t & \epsilon_0 \\  
                                  \end{array} \right) \label{h3} \\ 
           h^{(4)} &=& \left( \begin{array}{llll}
                               \epsilon_0 & t & t & 0 \\
                               t & \epsilon_0 & 0 & t \\
                               t & 0 & \epsilon_0 & t \\
                               0 & t & t & \epsilon_0 \\
                                  \end{array} \right)
         \end{eqnarray}
For the square lattice we choose $\epsilon_0=0$ and for the triangular lattice
$\epsilon_0=-0.83$, so that the Fermi level coincides with $\omega=0$. 

Instead of expressing the non-diagonal hybridization matrix $\Gamma(i\omega_n)$ 
in a site basis, it is convenient to go over to the molecular 
orbital basis in which $G_0(i\omega_n)$ is diagonal. Assuming that 
each component $G_{0,m}(i\omega_n)$ couples only with its own bath,
we have 
\begin{eqnarray}
 G_{0,m}(i\omega_n) &\approx&  G^{cl}_{0,m}(i\omega_n) \nonumber\\
    &=&    \left( i\omega_n + \mu -\epsilon_m -
           \sum_k  \frac{\vert V_{mk}\vert^2}{i\omega_n - \epsilon_k}\right)^{-1}
   \label{G0m}
\end{eqnarray}
where $\epsilon_m$ represents an impurity level, $\epsilon_k$ the bath
levels, and $V_{mk}$ the hybridization matrix elements. The incorporation 
of the impurity level $\epsilon_m$ ensures a much better fit of
$G_{0,m}(i\omega_n)$ than by projecting only onto bath orbitals.   

For instance, for $n_c=3$ and $n_s=12$ (i.e., three bath levels per impurity 
orbital), each component 
$G_{0,m}(i\omega_n)$ is fitted using seven parameters: one impurity level 
$\epsilon_m$, three bath levels $\epsilon_k$ and three hopping integrals
$V_{mk}$. Since according to Eq.~(\ref{g3}) there are two independent functions, 
we use a total of 14 fit parameters to represent these two $G_{0,m}$ components.   
This procedure allows for a considerably more flexible projection
of the Green's function matrix $G_0(i\omega_n)$ onto the bath. In a site basis
for an isotropic triangular lattice (taking again three bath levels per site) 
one would have instead only 6 fit parameters if each site couples to 
its own bath. Effectively, therefore, the molecular orbital basis accounts
for several additional cross hybridization terms as well as internal cluster
couplings (see below). Moreover, it is much more 
reliable to fit the two independent molecular orbital components 
$G_{0,m}(i\omega_n)$ than a non-diagonal site matrix $G_{0,ij}(i\omega_n)$ 
with an equivalent number of parameters. Analogous considerations hold
for the two-site and four-site clusters of the square lattice. 
For example, for $n_c=4$, $n_s=12$ there are two independent functions 
$G_{0,1}$ and $G_{0,2}$ ($G_{0,4}$ is related to $G_{0,1}$), giving a total 
of 10 fit parameters, compared to only two parameters in a simple site 
picture with fourforld and particle hole symmetry.

\begin{figure}[t]  
\begin{center}
\includegraphics[width=4.5cm,height=6.5cm,angle=-90]{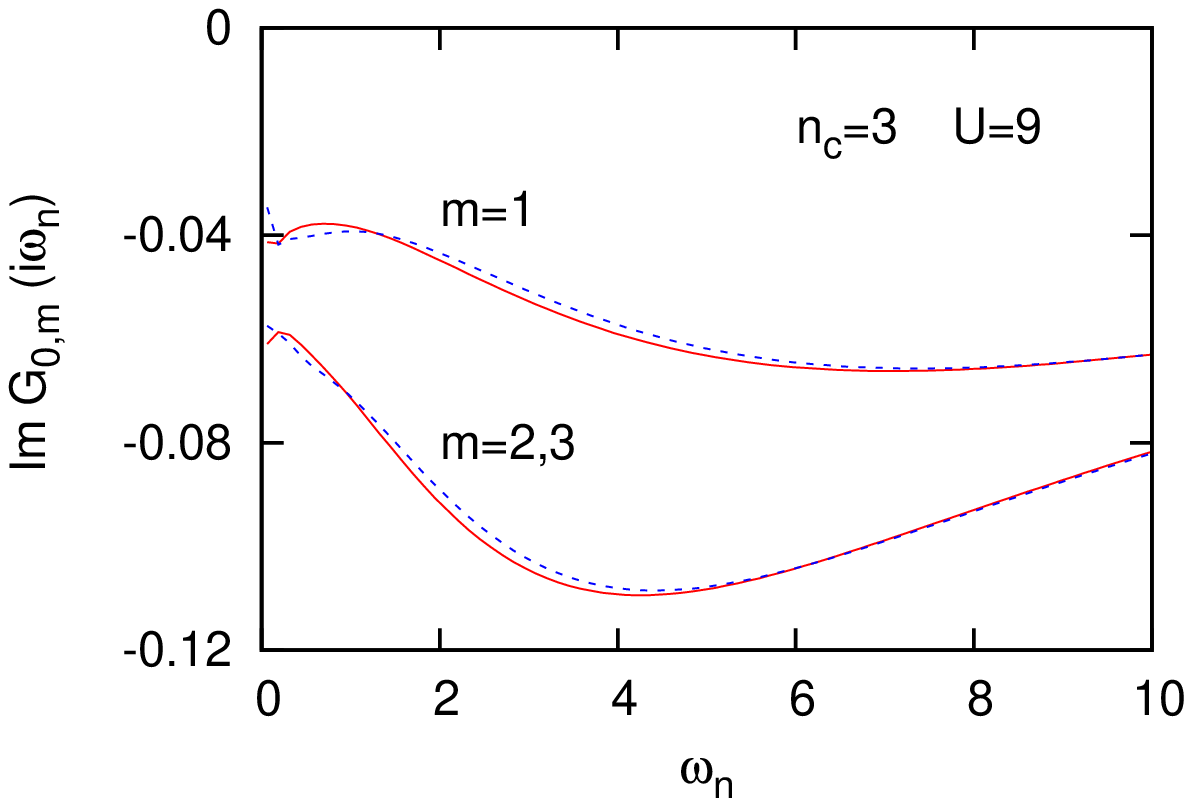}
\includegraphics[width=4.5cm,height=6.5cm,angle=-90]{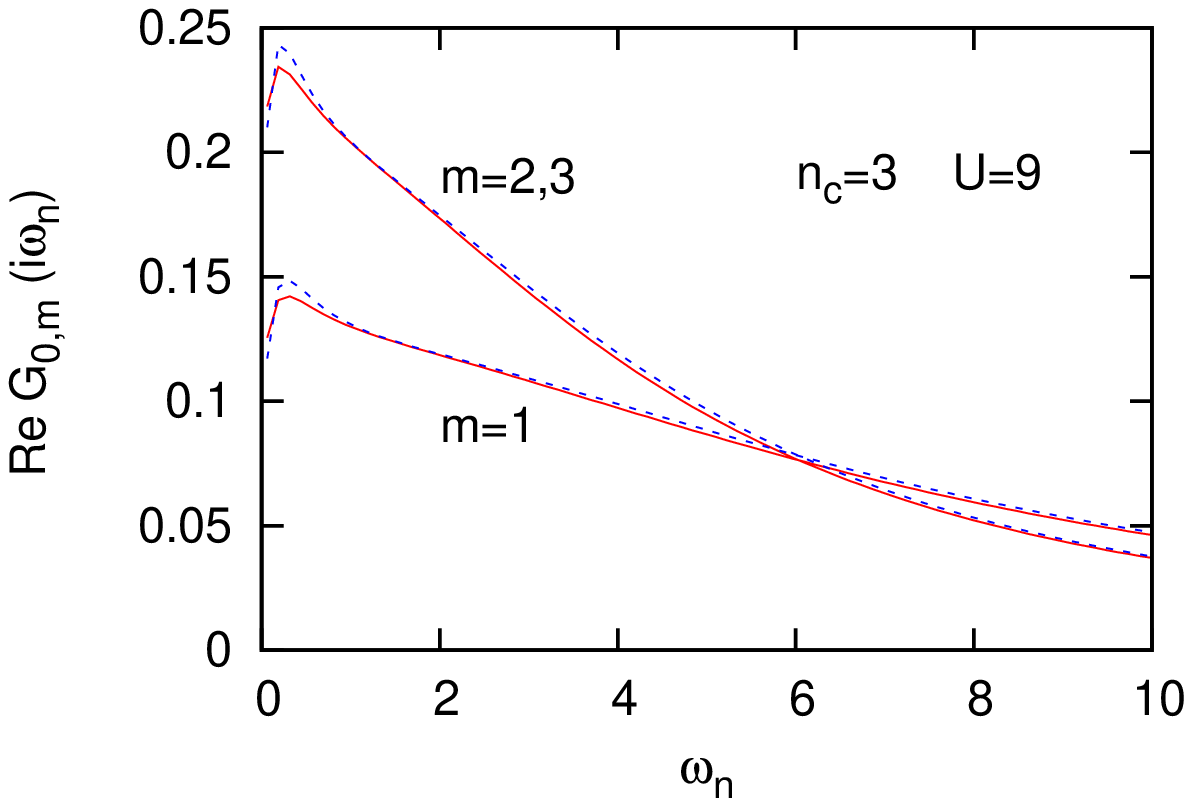}
\end{center} 
\caption{(Color online)
Projection of lattice Green's function components $G_{0,m}(i\omega_n)$ 
onto bath for $n_c=3$, $U=9$, $T=0.02$. Upper panel: Im\,$G_{0,m}$,
lower panel: Re\,$G_{0,m}$. 
Red curves: lattice Green's functions,  
blue curves: approximate expression, right-hand side of Eq.~(\ref{G0m}). 
}\end{figure}

Figure 2 illustrates the typical quality of the projection of the lattice 
components of $G_{0,m}(i\omega_n)$ onto the bath for $n_c=3$ and $n_s=9$.
Thus, although only two bath levels per orbital are included (i.e., using 
five parameters per orbital), the fit of both real and imaginary parts is 
excellent. For these cluster sizes and low temperatures, iterations take 
only a few minutes. Fits of similar quality are achieved for multi-orbital
materials.\cite{prb08,perroni,naco,tetra,v2o3-al} To achieve even better
agreement at low frequencies, it is preferable to minimize not the 
bare difference  $G_{0,m}(i\omega_n) - G^{cl}_{0,m}(i\omega_n)$ but to divide
these functions first by $\omega_n$.

We now discuss the evaluation of the finite temperature interacting cluster 
Green's function. If this step is carried out in the diagonal molecular 
orbital basis, the Coulomb interaction must be expressed as a matrix containing 
many inter-orbital components. For $n_c=4$ it can be easily shown that 
$U_{m_1m_2m_3m_4}=U/4$ for 64 of the possible 256 configurations. 
All other matrix elements vanish. This step can be circumvented by working 
in a mixed basis consisting of cluster sites $i$ and bath orbitals $k$. 
We illustrate this procedure here for the triangular lattice with $n_c=3$. 
Let us denote the transformation between sites and orbitals as $T^{(n_c)}$, 
where  
        \begin{eqnarray}
           T^{(3)}_{im} &=& \left( \begin{array}{rrr}
                               1/\sqrt3 & -2/\sqrt6 & 0 \\
                               1/\sqrt3 &  1/\sqrt6 & 1/\sqrt2 \\
                               1/\sqrt3 &  1/\sqrt6 &-1/\sqrt2 \\  
                                  \end{array} \right)  . 
         \end{eqnarray}
In this mixed basis, the effective site block of the cluster Hamiltonian 
becomes
        \begin{eqnarray}
           h^{(3)} &=& \left( \begin{array}{rrr}
                               \epsilon' & t'        & t' \\
                               t'        & \epsilon' & t' \\
                               t'        & t'        & \epsilon' \\  
                                  \end{array} \right) 
         \end{eqnarray}
with $\epsilon'=\epsilon_0 + (\epsilon_1+2\epsilon_2)/3$ and 
$t'=t + (\epsilon_1-\epsilon_2)/3$, where $\epsilon_0$ and $t$ are the 
elements of the original cluster Hamiltonian defined in Eq.~(\ref{h3}).
The new terms involving the molecular orbital cluster levels 
$\epsilon_m$ arise from the projection specified in  Eq.~(\ref{G0m}).     
In the mixed basis, the hybridization matrix elements $V_{mk}$ 
between cluster and bath orbitals introduced in Eq.~(\ref{G0m}) are 
transformed to new hybridization matrix elements between cluster sites 
$i$ and bath orbitals $k$. They are given by
\begin{equation}
       V'_{ik} = (T^{(3)} V)_{ik} = \sum_m T^{(3)}_{im} V_{mk}\ .
\end{equation}

Using the elements $\epsilon'$, $t'$ and $V'_{ik}$ together with the 
on-site Coulomb energy $U$, the non-diagonal interacting cluster Green's 
function at finite temperature is derived from the expression\cite{perroni,luca}
\begin{eqnarray}
 G^{cl}_{ij}(i\omega_n) &=& \frac{1}{Z} \sum_{\nu\mu}\,e^{-\beta E_\nu}\, 
          \Big(\frac{\langle\nu\vert c_{i\sigma}  \vert\mu\rangle 
                     \langle\mu\vert c_{j\sigma}^+\vert\nu\rangle}
                                  {E_\nu - E_\mu + i\omega_n}            \nonumber\\
       &&\hskip9mm + \ \ \frac{\langle\nu\vert c_{i\sigma}^+\vert\mu\rangle 
                               \langle\mu\vert c_{j\sigma}  \vert\nu\rangle}
                                  {E_\mu - E_\nu + i\omega_n} \Big)   
     \label{Gcl}
\end{eqnarray}
where $E_\nu$ and $|\nu \rangle$  denote the eigenvalues and eigenvectors of the 
impurity Hamiltonian, $\beta=1/k_B T$ and $Z=\sum_\nu {\rm exp}(-\beta E_\nu)$ 
is the partition function. At low temperatures only a relatively small number of 
excited states in few spin sectors contributes to $G^{cl}_{ij}$. They can be 
efficiently evaluated using the Arnoldi algorithm.\cite{arnoldi} The excited 
state Green's functions are computed using the Lanczos procedure. Further 
details can be found in Ref.\cite{perroni}. The non-diagonal elements
of $G^{cl}_{ij}$ are derived by first evaluating the diagonal components
$G^{cl}_{ii}$ and then using the relation\cite{senechal3} 
\begin{equation}
     G^{cl}_{(i+j)(i+j)} = G^{cl}_{ii}+G^{cl}_{ij}+G^{cl}_{ji}+ G^{cl}_{jj}.
\end{equation}
Since $G^{cl}_{ij}=G^{cl}_{ji}$, this yields: 
\begin{equation}
     G^{cl}_{ij} = \frac{1}{2}(G^{cl}_{(i+j)(i+j)} - G^{cl}_{ii} - G^{cl}_{jj}).
\end{equation}

For the two-site cluster,
we have used 3 or 4 bath levels for each impurity orbital ($n_s=8$ or 10),
for the three-site cluster 2 or 3 bath levels per impurity orbital
($n_s=9$ or 12), and for the four-site cluster 2 bath levels per impurity 
orbital ($n_s=12$). $G^{cl}_{ij}(i\omega_n)$ obeys the same symmetry properties 
as the lattice Green's functions given in Eqs.~(\ref{G2}--\ref{G4}). 
It therefore can be diagonalized as indicated in Eqs.~(\ref{g2}--\ref{g4}).
We denote these diagonal elements as   $G^{cl}_{m}(i\omega_n)$. For $n_c=4$,
we have checked that the evaluation of $G^{cl}_{m}(i\omega_n)$ in the 
non-diagonal site - orbital basis and in the diagonal molecular 
orbital basis yield identical results. 

The key assumption in DMFT is now that the resulting impurity cluster self-energy
is a physically reasonable representation of the lattice self-energy. Thus, 
using a relation analogous to Eq.~(\ref{G0}),  we find
\begin{eqnarray}
 \Sigma^{cl}_{m}(i\omega_n) &=&  1/G^{cl}_{0,m}(i\omega_n)-1/G^{cl}_{m}(i\omega_n) 
                                       \nonumber\\
        &\approx& \Sigma_{m}(i\omega_n)  .  
                                         \label{S}
\end{eqnarray}
After transforming $\Sigma_{m}(i\omega_n)$ back to the non-diagonal site basis,
it is used as input in the lattice Green's function Eq.~(\ref{G}) in the
next iteration step.  

To summarize the procedure discussed above, the multi-site ED/DMFT 
calculation consists of the following steps:\\
(a) evaluate the lattice Green's function $G_{ij}(i\omega_n)$, Eq.~(\ref{G}),
in the non-diagonal site basis, using as input  the self-energy obtained 
in a previous iteration step. The entire iteration 
procedure is started at small $U$ with $\Sigma=0$.\\
(b) transform $G_{ij}$ and $\Sigma_{ij}$ to the diagonal molecular
orbital basis and compute the components $G_{0,m}$. \\ 
(c) project the $G_{0,m}(i\omega_n)$ onto independent baths to determine 
$\epsilon_m$,  $\epsilon_k$ and $V_{mk}$ as indicated in Eq.~(\ref{G0m}).\\
(d) from the fit parameters $\epsilon_m$ and $V_{mk}$ determine the
Hamiltonian matrix elements $\epsilon'$, $t'$ and  $V'_{ik}$  
in the mixed site - orbital basis. \\
(e) evaluate the non-diagonal cluster Green's function 
$G^{cl}_{ij}(i\omega_n)$ using the Arnoldi and Lanczos methods. \\
(f) transform this Green's function to the diagonal orbital basis and 
compute the cluster self-energy components $\Sigma^{cl}_{m}(i\omega_n)$
defined in Eq.~(\ref{S}).\\

We emphasize that ED/DMFT involves, at each iteration, two projections:
(1) The lattice Green's function $G_0$ is projected onto the cluster
Green's function $G_0^{cl}$, as indicated in Eq.~(\ref{G0m}). 
By definition, $G_0$ has a continuous spectrum at real frequencies, 
while $G_0^{cl}$ is discrete.
(2) The cluster self-energy $\Sigma^{cl}$, which evidently has a
discrete spectrum at real $\omega$, is used as an approximation of the  
lattice self-energy $\Sigma$, which by definition is continuous along
the real frequency axis. 
Thus, both projections, $G_0\approx G_0^{cl}$ and $\Sigma^{cl}\approx\Sigma$, 
rely on the well-known fact that continuous and discrete spectra at real
$\omega$ can yield nearly identical distributions at Matsubara frequencies.
Since the cluster size determines the number of discrete spectral features
of $G_0^{cl}$ and $\Sigma^{cl}$, there exists evidently an infinite number 
of discrete spectra which may in principle be used to represent the    
continuous spectra of the lattice quantities $G_0$ and $\Sigma$. 

\section{Results and Discussion}

\begin{figure}[t]  
\begin{center}
\includegraphics[width=4.5cm,height=6.5cm,angle=-90]{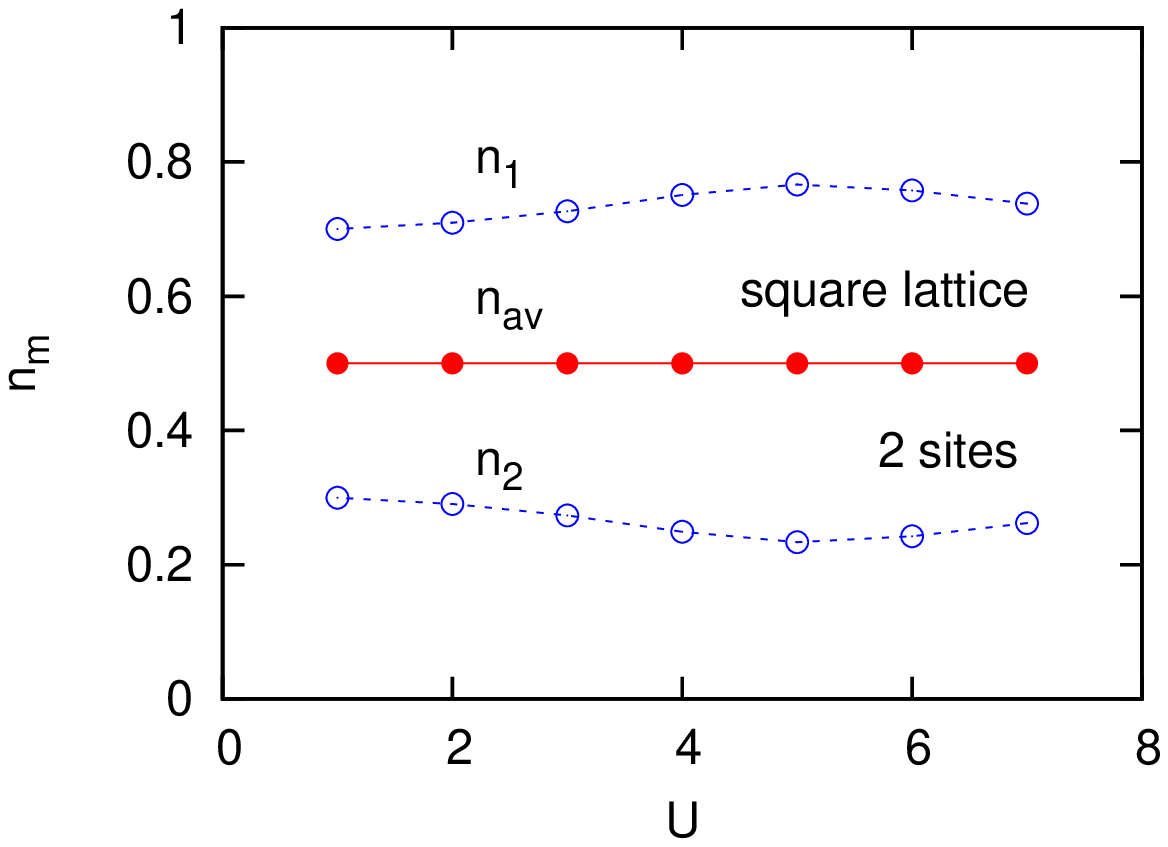}
\includegraphics[width=4.5cm,height=6.5cm,angle=-90]{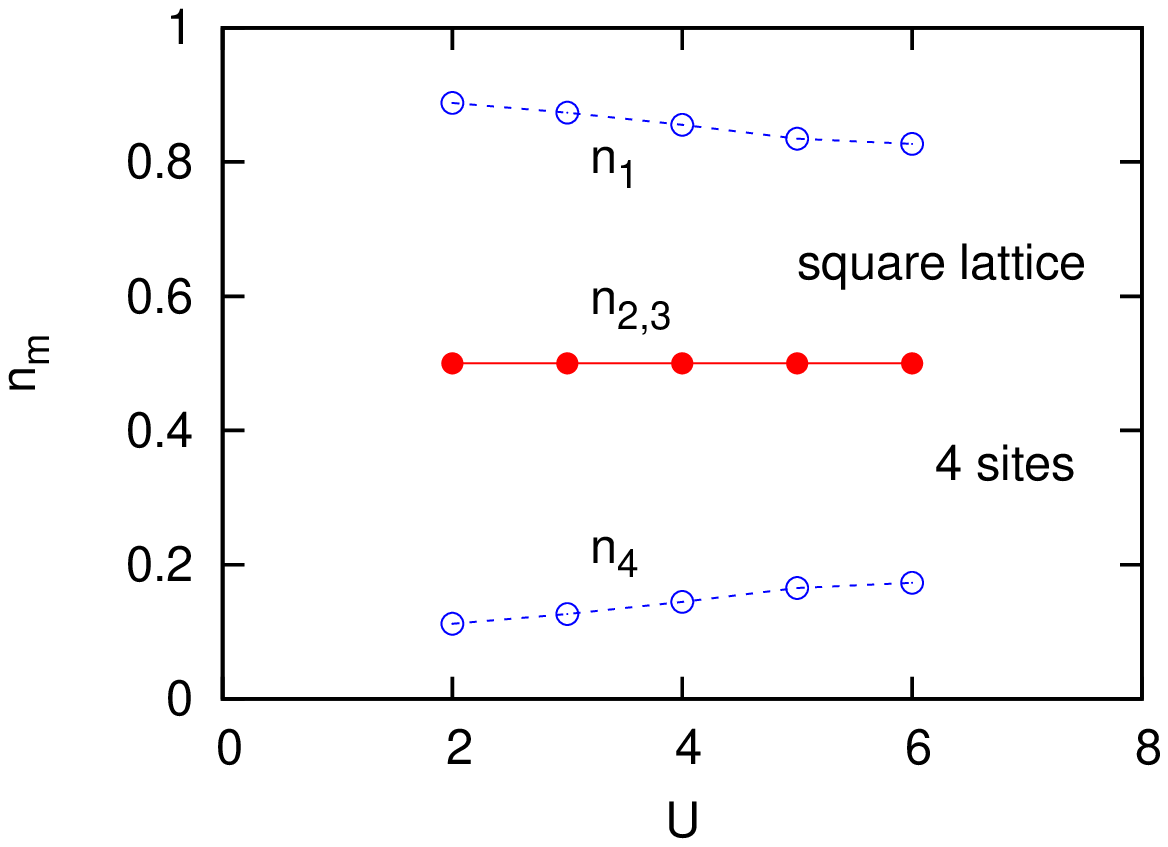}
\includegraphics[width=4.5cm,height=6.5cm,angle=-90]{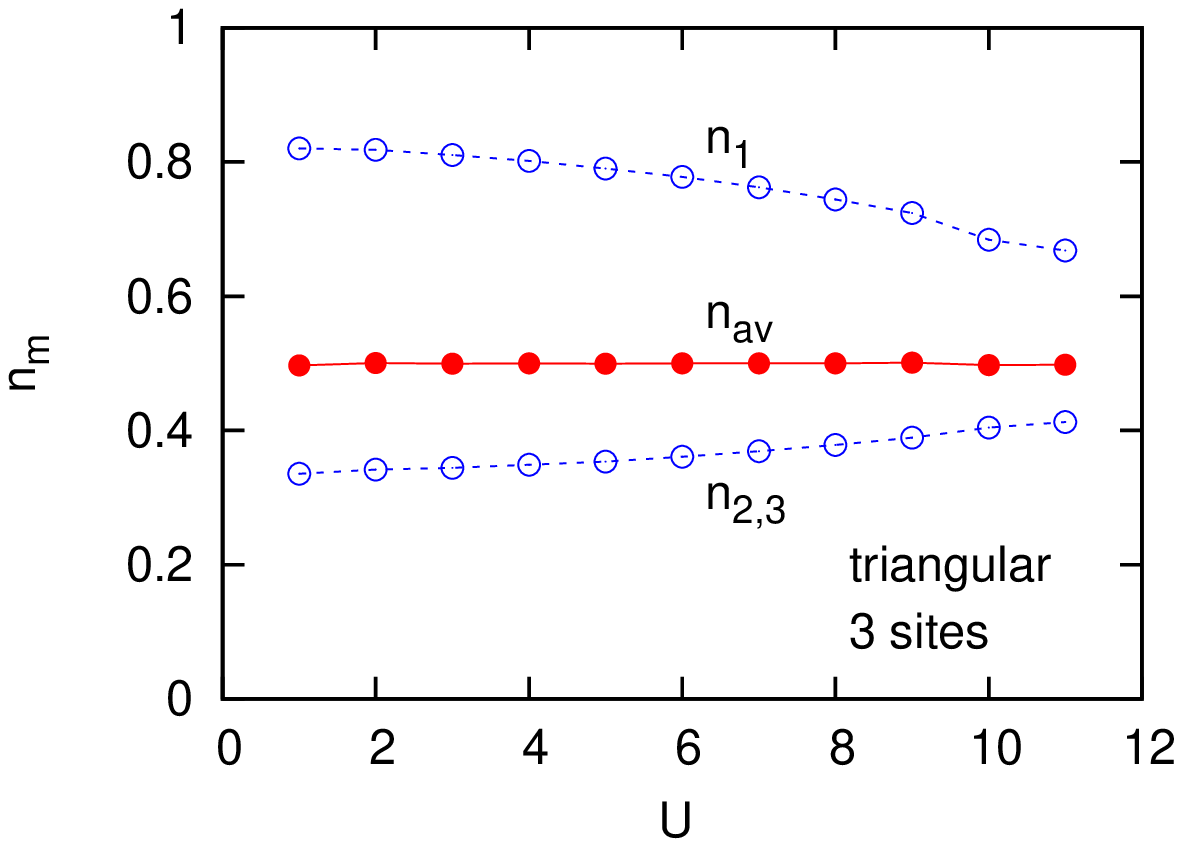}
\end{center}
\caption{(Color online)
Occupancies of cluster molecular orbitals (per spin) for two- and four-site
clusters of square lattice (top panels), and of three-site cluster of 
triangular lattice (bottom panel) as functions of Coulomb energy $U$.
The temperature is $T=0.02$. The sum of these molecular occupancies is 
$n_c/2$ and their average is $0.5$.
The Mott transition for increasing $U$ occurs at 
$U_{c2}\approx 5.5$ for the square lattice ($n_c=2$ and $n_c=4$), and near 
$U_{c2}\approx 9.5$ for the triangular lattice ($n_c=3$).  
}\end{figure}

Figure~3 shows the occupancies of the cluster molecular orbitals for three 
cluster sizes as functions of increasing $U$. For the square lattice 
the Mott transition occurs near $U_{c2}\approx 5.5$,
while for the triangular lattice $U_{c2}\approx 9.5$ (see below). 
Evidently all orbital occupancies vary smoothly across the transition.
There is no indication of orbital selective Mott transitions, nor for 
complete filling or emptying of any orbitals at large Coulomb energies.  
This behavior differs qualitatively from the one found
in materials such as LaTiO$_3$, V$_2$O$_3$, and Ca$_2$RuO$_4$, where 
the insulating phase exhibits nearly complete orbital polarization as 
a result of a Coulomb driven enhancement of the crystal field splitting 
between $t_{2g}$ orbitals\cite{pavarini,prb08,keller,poteryaev,anisimov,prl07} 
(see following section).

\begin{figure}[t]  
\begin{center}
\includegraphics[width=4.5cm,height=6.5cm,angle=-90]{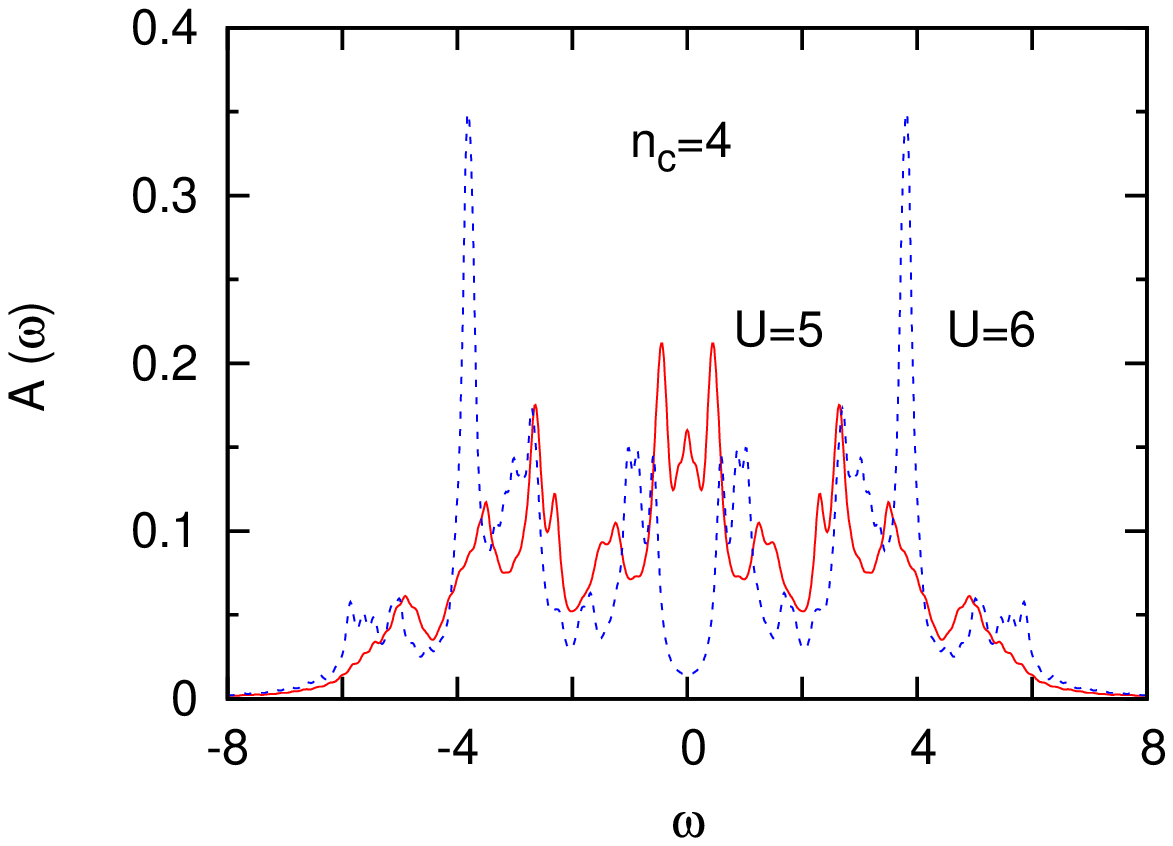}
\includegraphics[width=4.5cm,height=6.5cm,angle=-90]{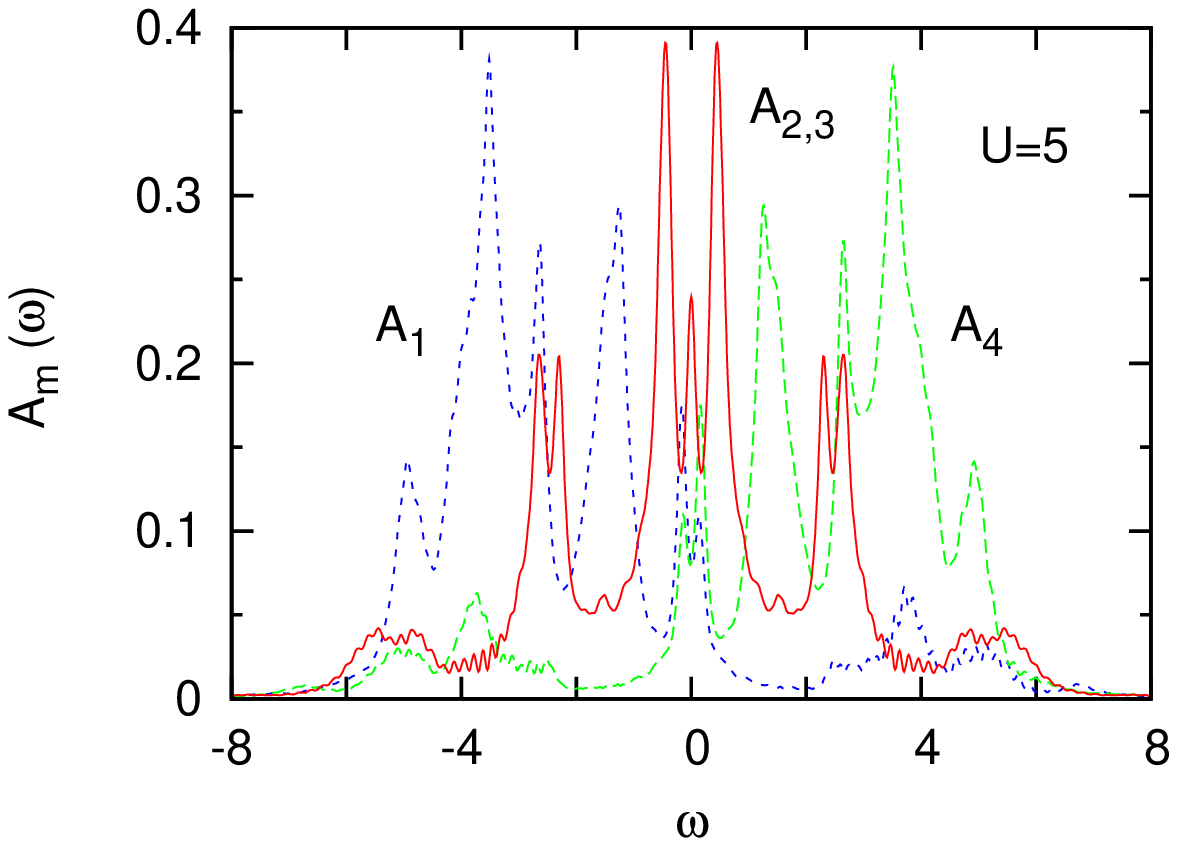}
\includegraphics[width=4.5cm,height=6.5cm,angle=-90]{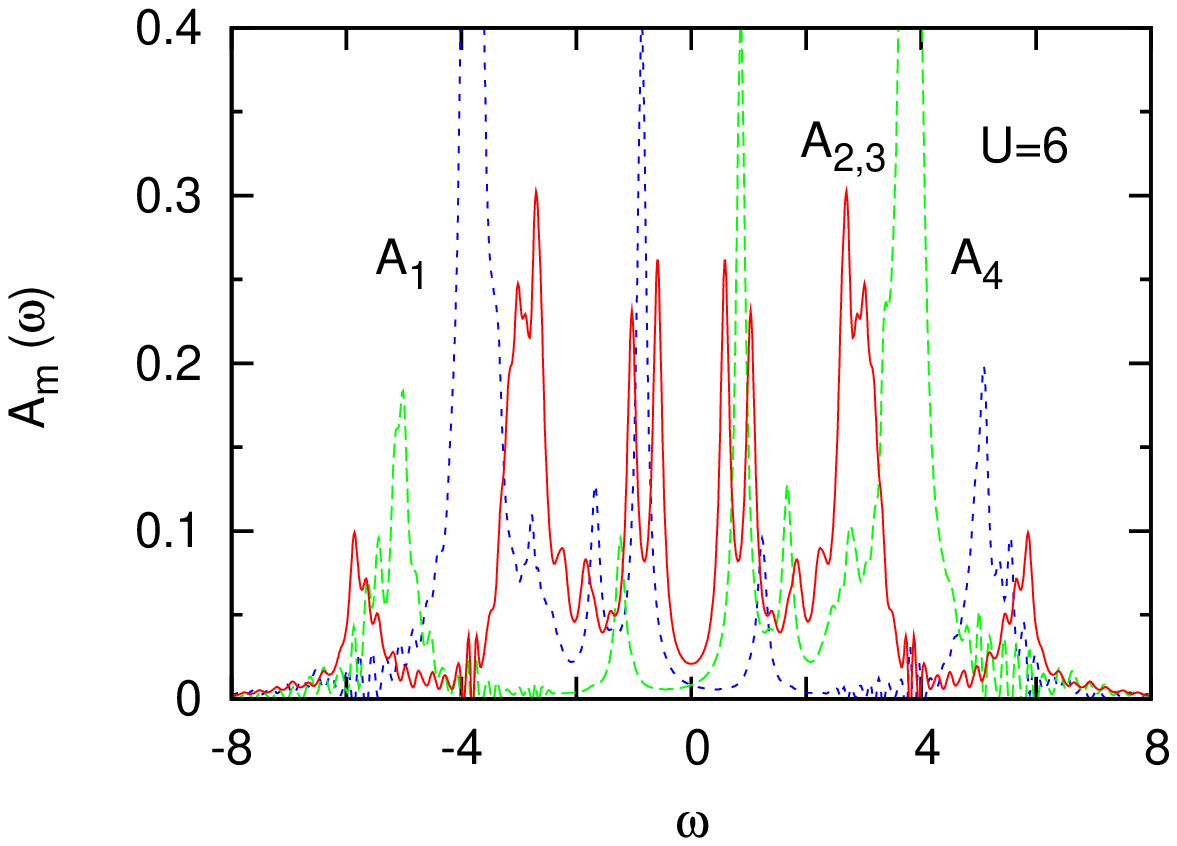}
\end{center}
\caption{(Color online)
Upper panel: total spectral distribution below and above the Mott transition 
($U_c\approx 5.5$) for the square lattice with $n_c=4$ at temperature 
$T=0.02$: $U=5$ (red curve) and $U=6$ (blue curve).
Lower two panels: molecular orbital contributions at $U=5$ and $U=6$; 
blue curves: $A_1(\omega)$  (associated with $\Gamma$), 
green curves: $A_4(\omega)$ ($M$), 
red curves: $A_{2,3}(\omega)$ ($X$) (broadening $\delta=0.1$).
}\end{figure}


The fact that all cluster orbitals remain partially occupied across the 
transition implies that the gap opens simultaneously in all orbitals.
This can be seen most clearly in the spectral distributions, as shown in 
Figure~4 for the square lattice with $n_c=4$. Since we are here concerned 
with the transition from metallic to insulating behavior, we show the 
spectra obtained from the interacting cluster Green's function, 
$A_{ij}(\omega)=-(1/\pi)\,{\rm Im}\, G^{cl}_{ij}(\omega+i\delta)$, 
Eq.~(\ref{Gcl}), with $\delta=0.1$. These spectra can be evaluated without 
requiring analytic continuation from Matsubara to real frequencies.
Using the transformations indicated in Eqs.~(\ref{g4}) and (\ref{rho4}), 
the cluster molecular orbital densities are given by   
$A_1    =A_{11}+2A_{12}+A_{14}$,
$A_2=A_3=A_{11}        -A_{14}$ and
$A_4    =A_{11}-2A_{12}+A_{14}$. 
The total density $A(\omega)=\sum_m A_m(\omega)/4$ coincides with the on-site 
distribution $A_{11}(\omega)$, which also represents the density corresponding 
to $M/2=(\pi/2,\pi/2)$ (see below). To our knowledge, this orbital decomposition 
has not been addressed before.  
Clearly, all orbitals contribute to the spectral weight at $E_F$ in the metallic 
phase, as well as to the lower and upper Hubbard bands in the insulating phase. 
The spectral distributions shown in the upper panel are consistent with those 
by Kyung et al.\cite{kyung2} and Zhang and Imada\cite{zhang} within ED/DMFT at 
$T=0$. The spectra reveal a characteristic four-peak structure, consisting of 
low-frequency peaks limiting the pseudogap due to short-range correlations and 
high-frequency peaks associated with the Hubbard bands.\cite{preuss,moreo,kyung2}
The small peak at $E_F$ in the metallic phase at $U=5$ appears only at finite $T$.
It vanishes at $T=0$. At such low frequencies, however, ED finite-size effects 
cannot be ruled out.   


\begin{figure}[t]  
\begin{center}
\includegraphics[width=4.5cm,height=6.5cm,angle=-90]{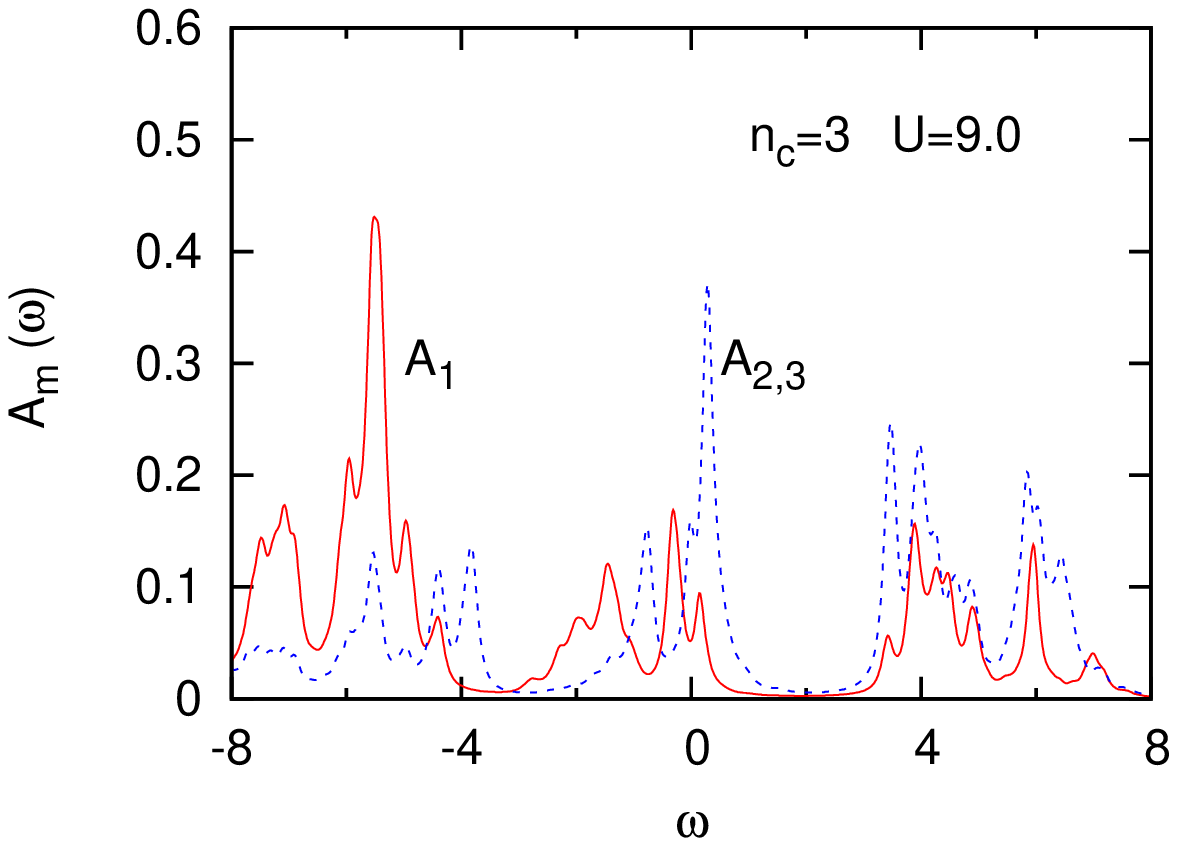}
\includegraphics[width=4.5cm,height=6.5cm,angle=-90]{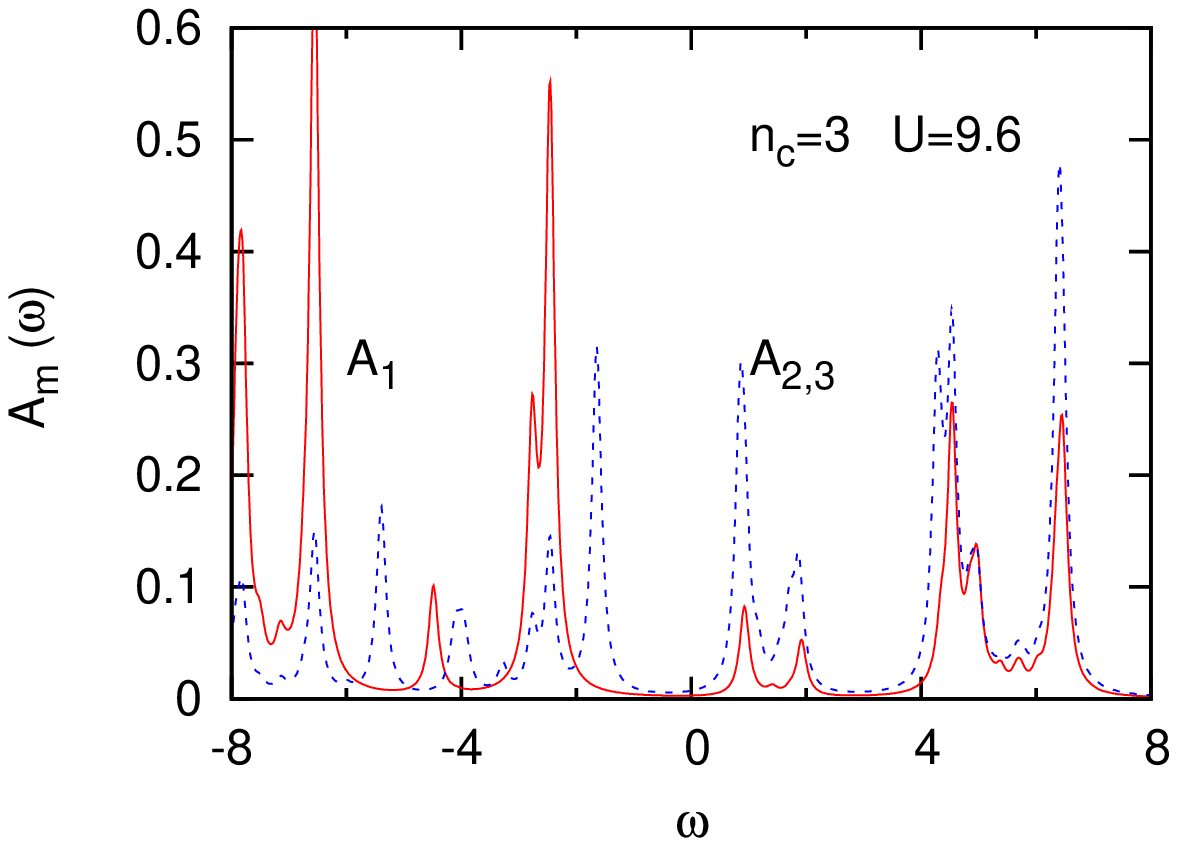}
\end{center}
\caption{(Color online)
Spectral distributions of cluster molecular orbitals below and above 
the Mott transition ($U_c\approx 9.5$) for the triangular lattice 
with $n_c=3$ at temperature $T=0.02$.
Upper panel: $U=9.0$; lower panel: $U=9.6$.
Blue curves: $A_1(\omega)$, red curves: $A_{2,3}(\omega)$.      
}\end{figure}

Similar results are obtained for the square lattice in the two-site cluster
model. Results for the triangular lattice with $n_c=3$ are shown in Figure~5. 
According to Eqs.~(\ref{g3}) and (\ref{rho3}) the cluster molecular orbital 
densities are given by $A_{1}=A_{11}+2A_{12}$ and $A_2=A_3=A_{11}-A_{12}$. 
As for the square lattice, the Mott gap opens simultaneously in all orbitals 
at about the same critical $U$, and all orbitals contribute to the lower and 
upper Hubbard bands.  
While the metallic phase of the unfrustrated square lattice close to the 
transition exhibits a pseudogap due to short-range antiferromagnetic 
correlations,\cite{kyung2,zhang,park,gull} this phenomenon is absent in the 
triangular lattice as a result of geometrical frustration.\cite{kyung3} 
  
\begin{figure}[t]  
\begin{center}
\includegraphics[width=4.5cm,height=6.5cm,angle=-90]{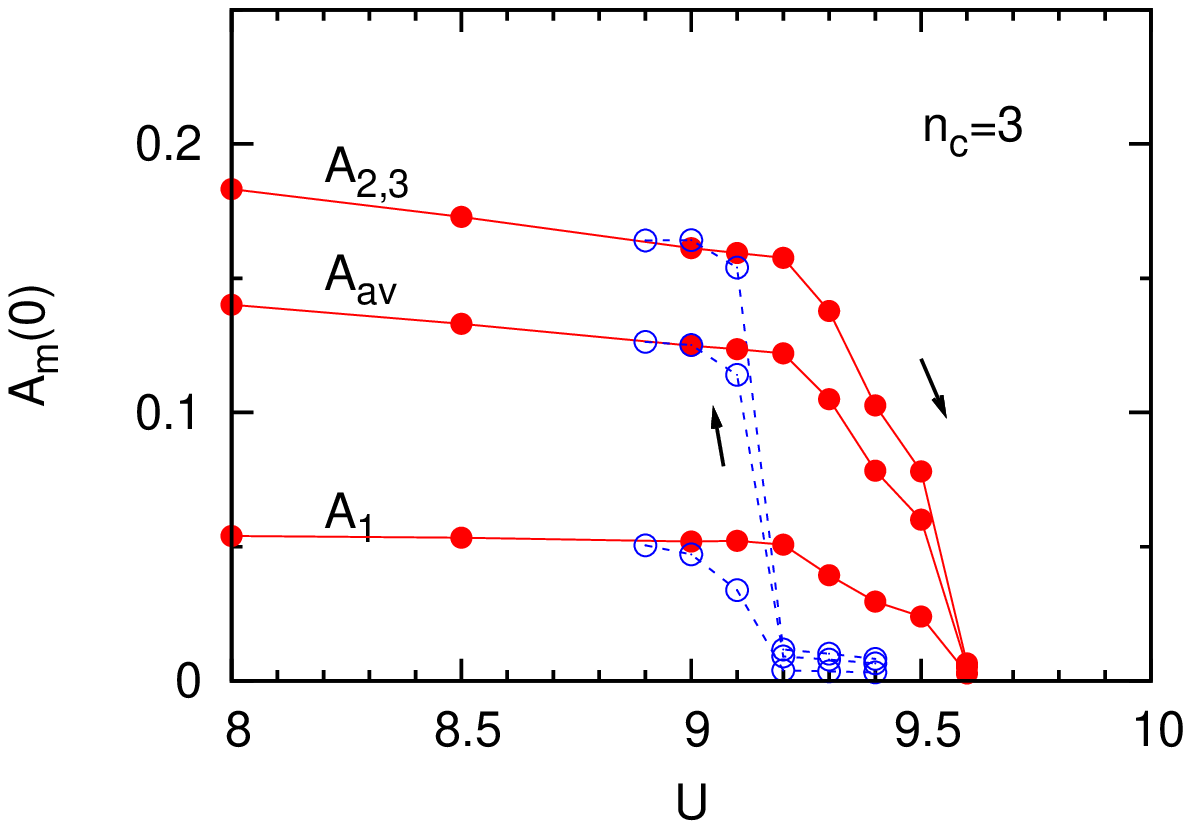}
\includegraphics[width=4.5cm,height=6.5cm,angle=-90]{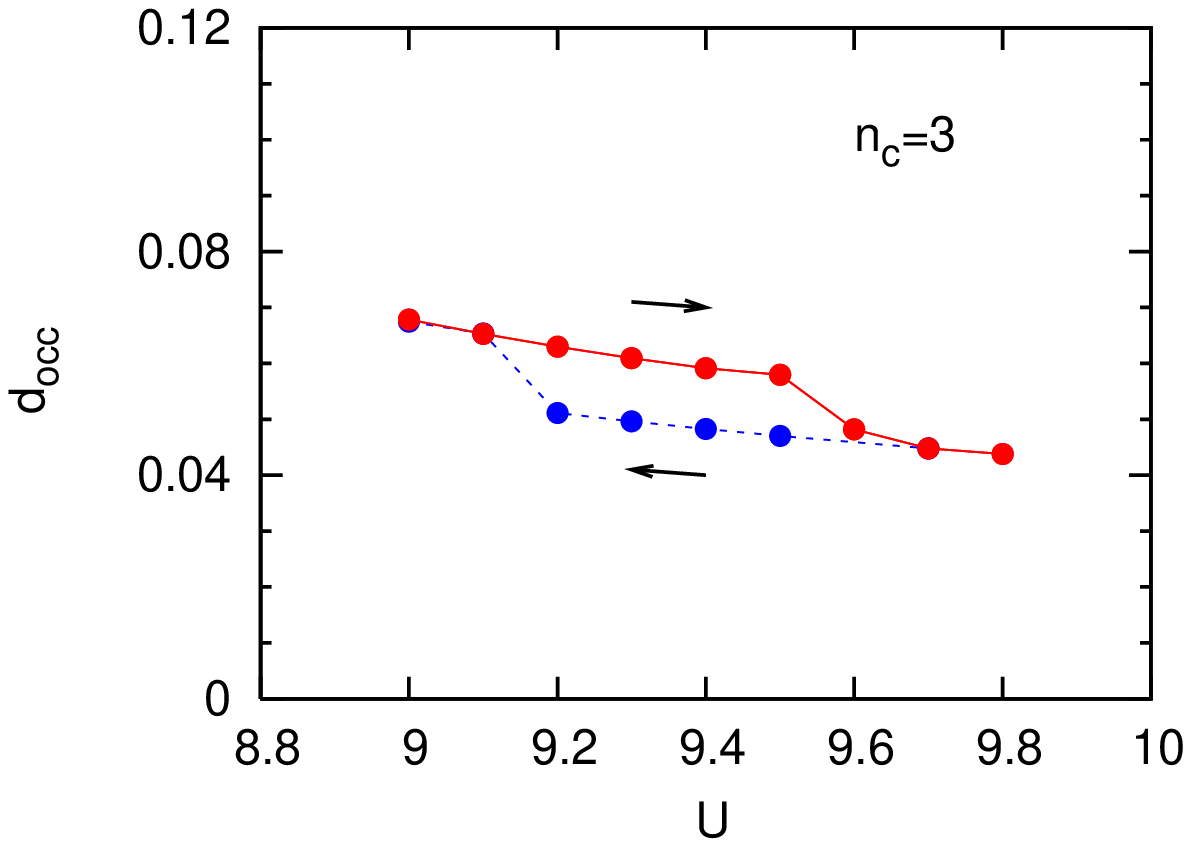}
\end{center}
\caption{(Color online)
Hysteresis behavior of spectral weights $A_m(0)$ of cluster molecular orbitals 
at $E_F=0$ (upper panel) and average double occupancy (lower panel) as functions 
of Coulomb energy for triangular lattice with $n_c=3$ ($T=0.02$). 
Red curves: increasing $U$, blue curves: decreasing $U$.
}\end{figure}

To illustrate the first-order nature of the metal insulator transition we show 
in the upper panel of Figure~6 the spectral weight of the $n_c=3$ cluster orbitals 
at $E_F=0$ as a function of $U$. The lower panel shows the average double occupancy 
$d_{\rm occ}=\sum_m \langle n_{m\uparrow}n_{m\downarrow}\rangle/3$. 
Both quantities exhibit hysteresis for increasing and decreasing $U$, indicating 
the coexistence of metallic and insulating solutions. 
The complete $T / U$ phase diagram will be published elsewhere. 
The phase diagram for the isotropic square lattice with $n_c=4$ was recently 
mapped out in detail by Park et al.\cite{park} 

The cluster molecular orbital components of the self-energy and Green's function 
may be used to derive an approximate expression for the momentum variation of the 
lattice self-energy and Green's function in the original Brillouin Zone:\cite{parcollet}
\begin{equation}
 \Sigma(\vec k,i\omega_n) \approx \frac{1}{n_c}
            \sum_{ij} e^{i\vec k\cdot(\vec R_i-\vec R_j)} \Sigma_{ij}(i\omega_n)
    \label{Slattice}
\end{equation}
where $\vec R_i$ are the cluster site positions and $\Sigma_{ij}$ are the site
components of the self-energy. An analogous expression holds for the lattice Green's 
function. For the clusters discussed above this superposition implies that the Mott 
gap opens uniformly along the Fermi surface since all orbitals undergo a common 
transition. 

Writing the site elements in terms of the orbital components, one has for $n_c=4$:
$\Sigma(\Gamma,i\omega_n)=\Sigma_1(i\omega_n)$,
$\Sigma(X     ,i\omega_n)=\Sigma_2(i\omega_n)=\Sigma_3(i\omega_n)$ and
$\Sigma(M     ,i\omega_n)=\Sigma_4(i\omega_n)$. 
In agreement with results of previous authors
\cite{parcollet,civelli,zhang,park,gull,ferrero,balzer} 
we find the behavior of $\Sigma(\Gamma,i\omega_n)$ and $\Sigma(M,i\omega_n)$ 
near the Mott transition to differ qualitatively from that of $\Sigma(X,i\omega_n)$ 
(not shown here): Whereas Im\,$\Sigma(X,i\omega_n)$ exhibits $\sim\omega_n$
variation at low frequencies in the metallic phase and $\sim 1/\omega_n$ variation
in the insulating phase (the real part vanishes because of particle hole symmetry,
see Fig.~1), Im\,$\Sigma(\Gamma,i\omega_n)$ and Im\,$\Sigma(M,i\omega_n)$ remain 
$\sim\omega_n$ in both phases, but their real parts increase rapidly across the 
transition. 

Although this behavior might suggest a Mott transition for the $X$ cluster orbitals
combined with a band-filling or band-emptying mechanism for the $\Gamma$ and $M$ 
orbitals, the orbital occupancies (Fig.~3) and spectral distributions (Fig.~4) 
demonstrate this not to be the case. Moreover, for the crucial question of whether 
or not the Mott gap opens simultaneously across the Fermi surface, it is important
to compare the self-energy at $X$ with its behavior at $M/2$, where the electron 
band also crosses $E_F$. According to Eq.~(\ref{Slattice}), 
$\Sigma(M/2,i\omega_n)$ coincides with the diagonal on-site element of the cluster 
self-energy, which is identical with the local lattice self-energy. Thus:
\begin{eqnarray}
     \Sigma(M/2,i\omega_n) &=& \frac{1}{4}[\Sigma(\Gamma,i\omega_n)+\Sigma(M,i\omega_n)
               \nonumber\\
          && \ \ \  + \          2\,  \Sigma(X,i\omega_n)].
\end{eqnarray}
The real parts of the first terms on the {\it rhs} cancel since the corresponding 
density of states components are mirrors of each other (see Fig.~1). Thus, 
$\Sigma(M/2,i\omega_n)$ 
and $\Sigma(X,i\omega_n)$ are purely imaginary because of particle hole symmetry. 
The above relation demonstrates that Coulomb correlations at the cold spot $M/2$ 
are essentially driven by those at the hot spot $X$. In fact, the magnitude of 
the self-energy at $M/2$ is a factor of 2 smaller than the singular term at $X$, 
with weak additional, non-singular contributions associated with $\Gamma$ and $M$. 

Along the Fermi surface between $X$ and $M/2$  (i.e. for $k_x+k_y=\pi$) 
the self-energy is given by
\begin{equation}
     \Sigma(\vec k,i\omega_n) = {\rm cos}^2(k_x)\Sigma(X,i\omega_n)
                               +{\rm sin}^2(k_x)\Sigma(M/2,i\omega_n). \\
   \label{XtoM/2}
\end{equation}
This function is imaginary, i.e., there are no band shifts due to a finite real 
part of the self-energy. Thus, within the four-site cluster DMFT, the opening 
of the Mott gap on the entire Fermi surface is determined solely by the singularity 
of Im\,$\Sigma(X,i\omega_n)$. 

The above analysis leads to a surprisingly simple picture for the momentum 
variation of correlations along the Fermi surface. It consists of two sinusoidal 
contributions: The singular $X$ term oscillates with amplitude $1$ at $X$ and 
$1/2$ at $M/2$, and the non-singular term due to the $\Gamma,\ M$ orbitals 
oscillates with amplitude $1$ at $M/2$ and zero at $X$.

\begin{figure}[t]  
\begin{center}
\includegraphics[width=4.5cm,height=6.5cm,angle=-90]{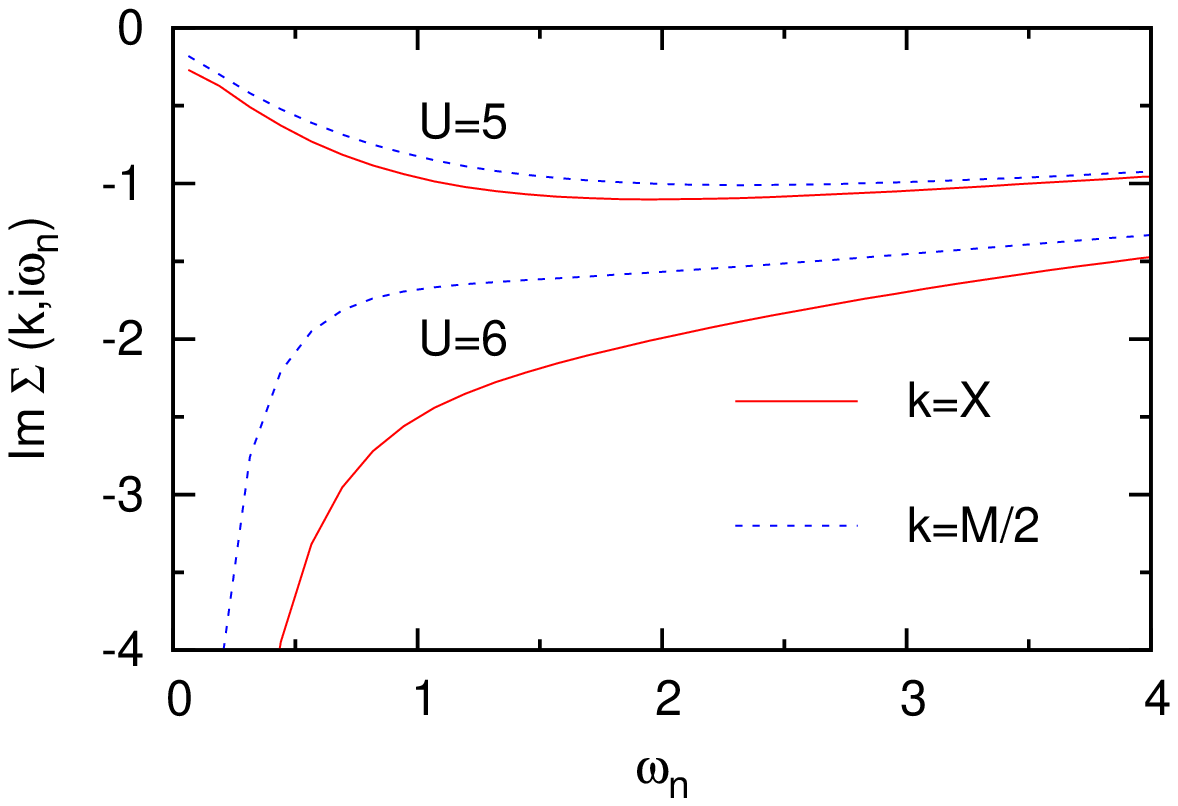}
\includegraphics[width=4.5cm,height=6.5cm,angle=-90]{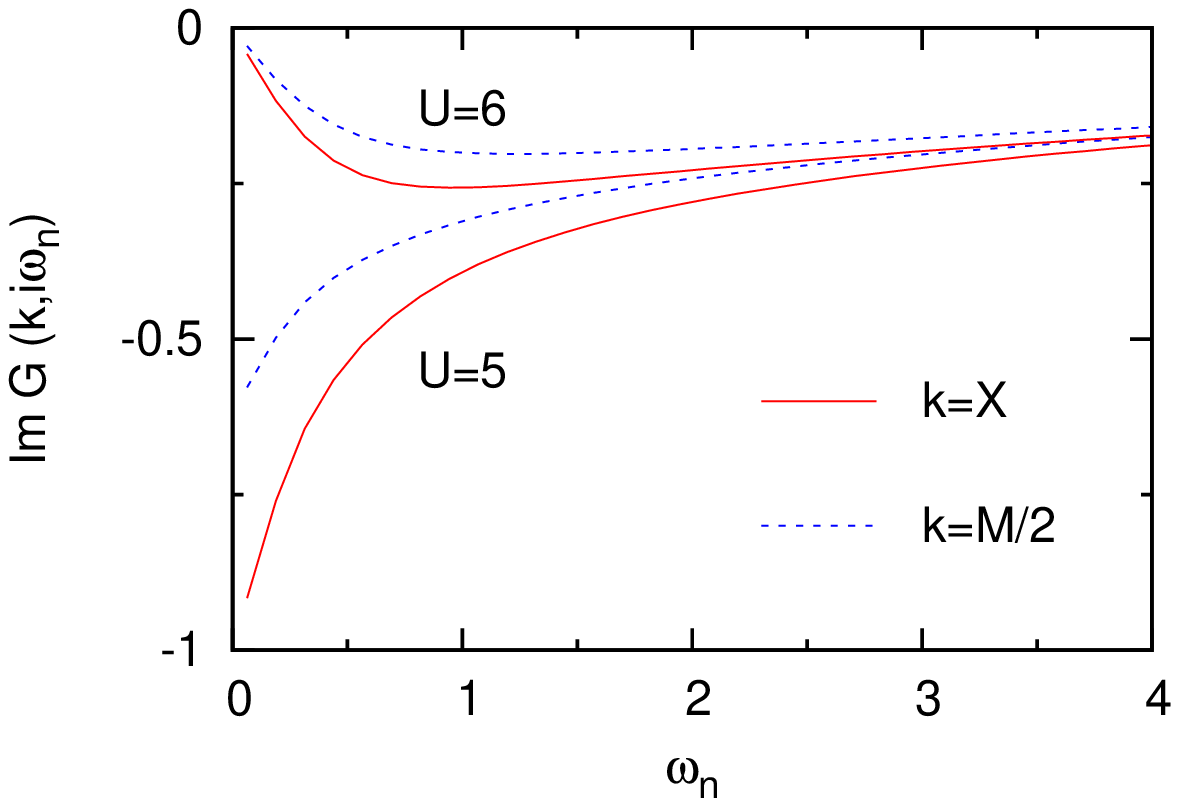}
\end{center}
\caption{(Color online)
Lattice self-energy, Eq.~(\ref{Slattice}), (upper panel) and Green's function 
(lower panel) at $X=(\pi,0)$ (red curves) and $M/2=(\pi/2,\pi/2)$ (blue curves) 
for metallic phase ($U=5$) and insulation phase ($U=6$) of square lattice 
($n_c=4$) at $T=0.02$. Both $\Sigma(\vec k,i\omega_n)$ and $G(\vec k,i\omega_n)$ 
are purely imaginary between $X$ and $M/2$. The sinusoidal variation of the 
self-energy between these points is given by Eq.~(\ref{XtoM/2}).
}\end{figure}

\begin{figure}[t]  
\begin{center}
\includegraphics[width=4.5cm,height=6.5cm,angle=-90]{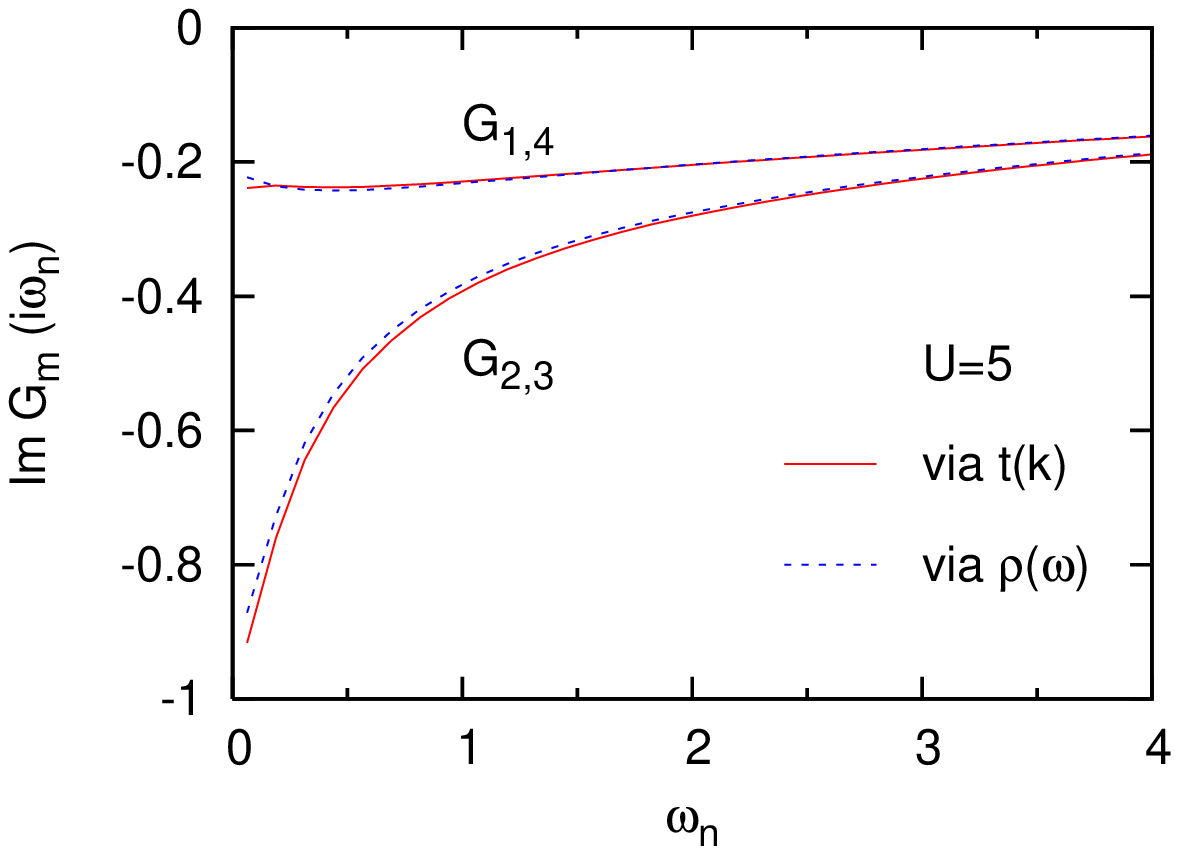}
\includegraphics[width=4.5cm,height=6.5cm,angle=-90]{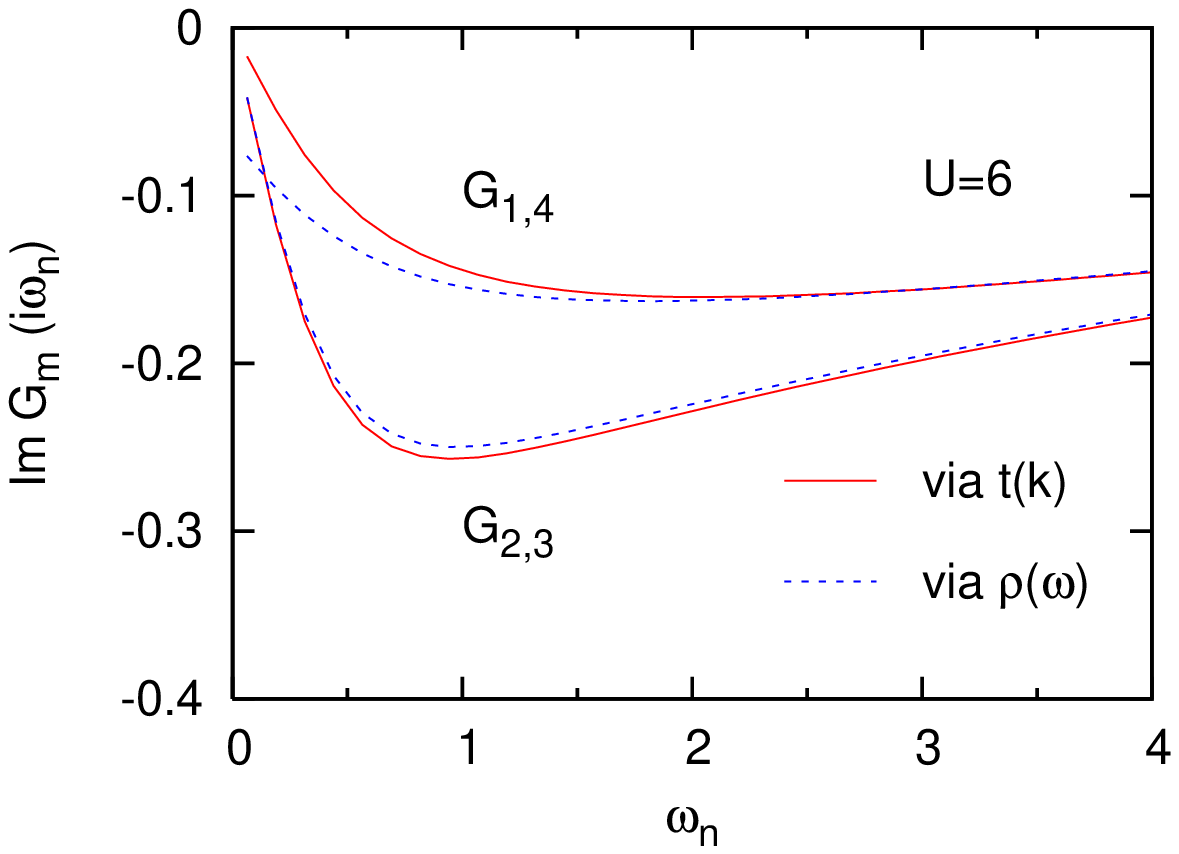}
\end{center}
\caption{(Color online)
Lattice Green's function Eq.~(2) in $n_c=4$ cluster molecular orbital basis 
for $U=5$ (upper panel) and $U=6$ (lower panel) at $T=0.02$.
Red curves: diagonal version of Eq.~(\ref{G});
blue curves: approximate expression, Eq.~(\ref{Grho}). 
}\end{figure}

This is illustrated in Figure 7 which shows the lattice self-energy and 
Green's function at $X$ and $M/2$ below and above the Mott transition.  
In the metallic phase, the self-energy at $X$ is seen to be 
larger than the one at $M/2$. Thus, quasi-particle lifetimes decrease between
$M/2$ and $X$. Nevertheless, the singular behavior at $M/2$ is governed 
by the one at $X$. Accordingly, the Green's function at $X$ and $M/2$ displays 
a change from metallic to insulating behavior at the same Coulomb energy.   

To understand the strong coupling between different sections of the 
Brillouin Zone it is important to recall that, in the diagonal cluster 
molecular orbital basis, the single-particle part of the lattice Hamiltonian 
appearing in Eq.~(\ref{G}) is not diagonal. Thus, the orbital component of the
lattice Green's function  
$G_m(i\omega_n)$ is influenced not only by the corresponding self-energy 
$\Sigma_m(i\omega_n)$, but by the other orbital elements as well. This point 
becomes clear if we compare $G_m(i\omega_n)$ with the approximation 
\begin{equation}
    G_m(i\omega_n) \approx \int d\omega\frac{\rho_m(\omega)}
              { i\omega_n + \mu - \omega - \Sigma_m(i\omega_n)}.
   \label{Grho}
\end{equation}
Figure 8 shows that in the metallic region at $U=5$ there is little difference
with regard to the actual $G_m(i\omega_n)$. Also, at $U=6$ the key components
responsible for the metal insulator transition, namely $G_{2,3}(i\omega_n)$ 
corresponding to $X=(\pi,0)$, are well represented by this approximation.
The $\Gamma,M$ components, $G_{1,4}(i\omega_n)$, however, do not reveal 
insulating behavior since at small $\omega_n$ the imaginary parts do not
extrapolate to zero. This demonstrates that the Mott gaps seen in 
$A_{1,4}(\omega)$ in Figs.~4 and 5 are not caused by the rapidly varying 
real parts of $\Sigma_{1,4}(i\omega_n)$. Instead, the gaps at $\Gamma$ and 
$M$ are driven by the singular behavior of Im\,$\Sigma_{2,3}(i\omega_n)$ 
which contributes to $A_{1,4}(\omega)$ via the non-diagonal elements of 
$t(\vec k)$.\cite{kmixing}
   
If the approximate components obtained via Eq.~(\ref{Grho}) were used
to generate the spectral distributions $A_m(\omega)$, it is clear that
only $A_{2,3}(\omega)$ corresponding to $X$ would exhibit the Mott transition, 
while $A_{1,4}(\omega)$ would retain considerable metallicity. This implies 
that the physics at the cold spot $M/2$ is incorrectly represented 
via Eq.~(\ref{Grho}), suggesting that the gap at $M/2$ does not open at
the same $U$ as at $X$. Instead, as argued above, the metal insulator
transition at $M/2$ is caused by the same self-energy terms as at $X$,
i.e., the Mott gap opens uniformly.
 
As pointed out earlier, in the molecular orbital basis the Coulomb matrix 
has a large number of non-zero elements. Thus, in this basis there is not 
only single-particle hybridization arising from $t(\vec k)$, but also 
strong inter-orbital Coulomb repulsion. Nevertheless, the above analysis
reveals that, although these Coulomb interaction terms are properly taken 
into account, the spectral distributions at $\Gamma$, $M$ and $M/2$ do not
exhibit a Mott gap unless the non-diagonal terms of $t(\vec k)$ in the
orbital basis are included. It is therefore the single-particle part of
the Hamiltonian that provides the correct connection between the self-energy
components and thereby generates the true momentum variation of the spectral 
distribution of the single band. 

Because of the strong coupling between orbitals, the notion that some of 
these orbitals undergo a Mott transition while others do not, does not appear 
appropriate. As shown consistently by all orbital-resolved spectra in the 
present multi-site ED/DMFT study, there is a single Mott transition common 
to all cluster molecular orbitals, implying a simultaneous opening of the 
Mott gap across the entire Fermi-surface. Nevertheless, in agreement with 
previous authors, we find pronounced momentum variation of quasi-particle
properties close to the Mott transition. 

In view of the approximate nature of the momentum variation of the lattice
self-energy and Green's function derived within the CDMFT it would be very
interesting to compare the above results with analogous ones obtained within 
the DCA,\cite{hettler,maier} in particular, since the cluster molecular orbital
components of the density of states, as stated above, differ appreciably between 
these two cluster DMFT schemes.\cite{haule} This comparison will be addressed 
in a future publication. 

The scenario discussed in this section differs strikingly from the one found 
for several multi-orbital materials which have been studied previously by 
various groups. For the sake of comparison we review some of these systems 
in the following section.  

\section{Comparison with multi-orbital systems}

\begin{figure}[t]  
\begin{center}
\includegraphics[width=4.5cm,height=6.5cm,angle=-90]{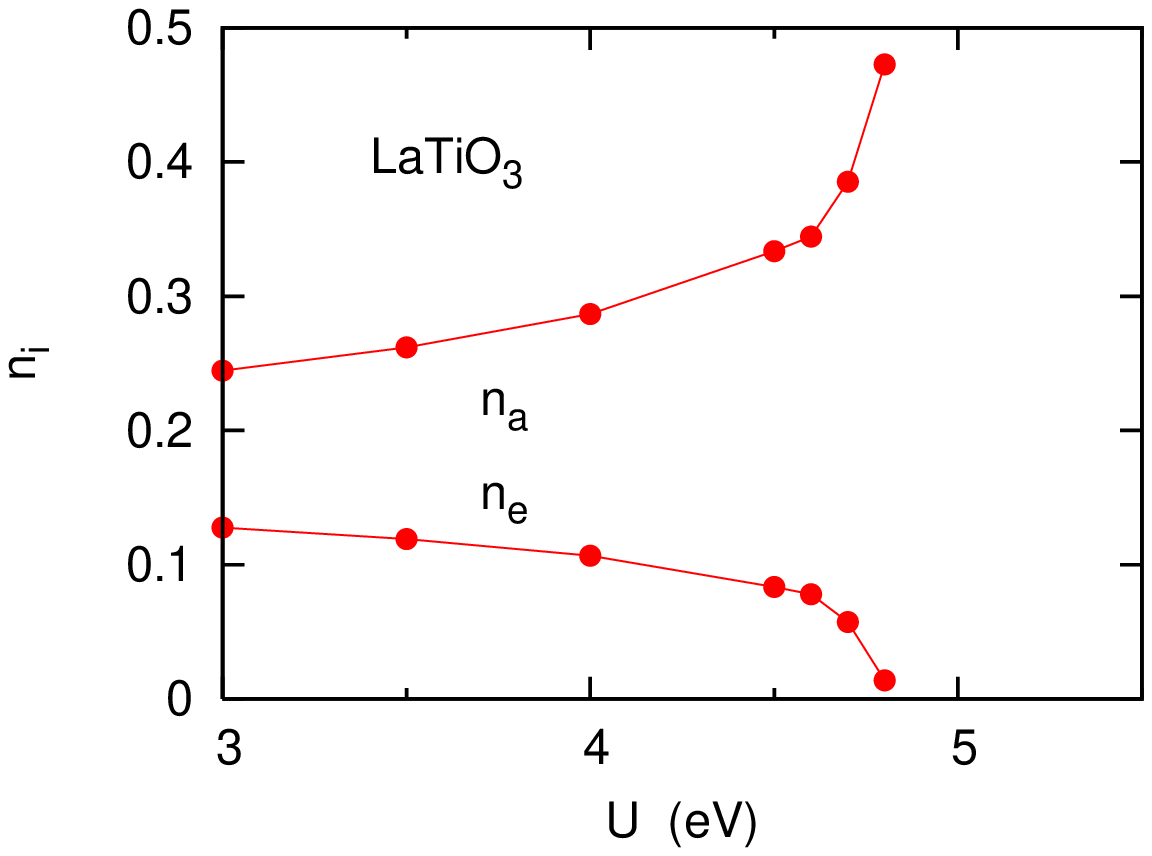}
\includegraphics[width=4.5cm,height=6.5cm,angle=-90]{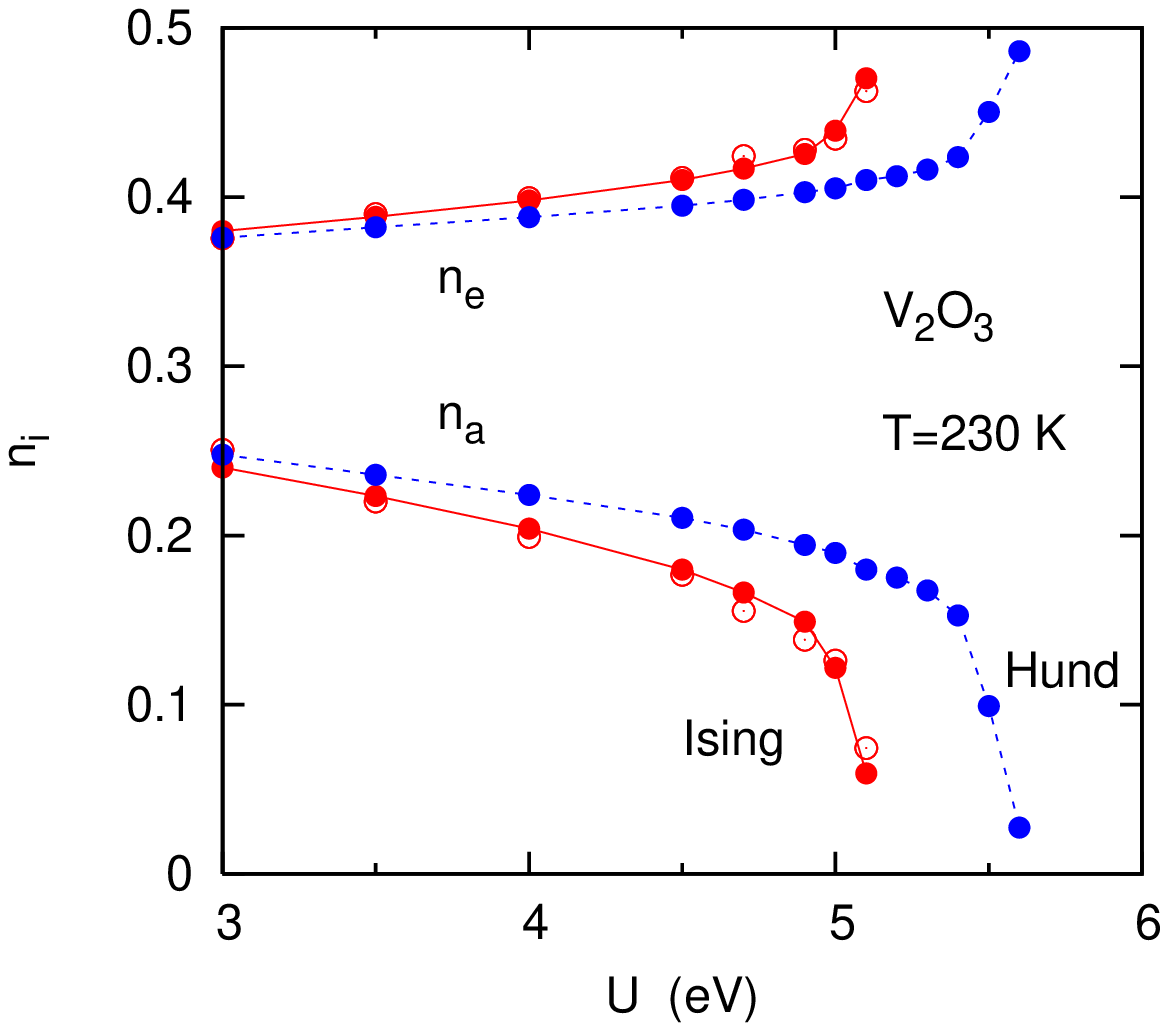}
\includegraphics[width=4.5cm,height=6.5cm,angle=-90]{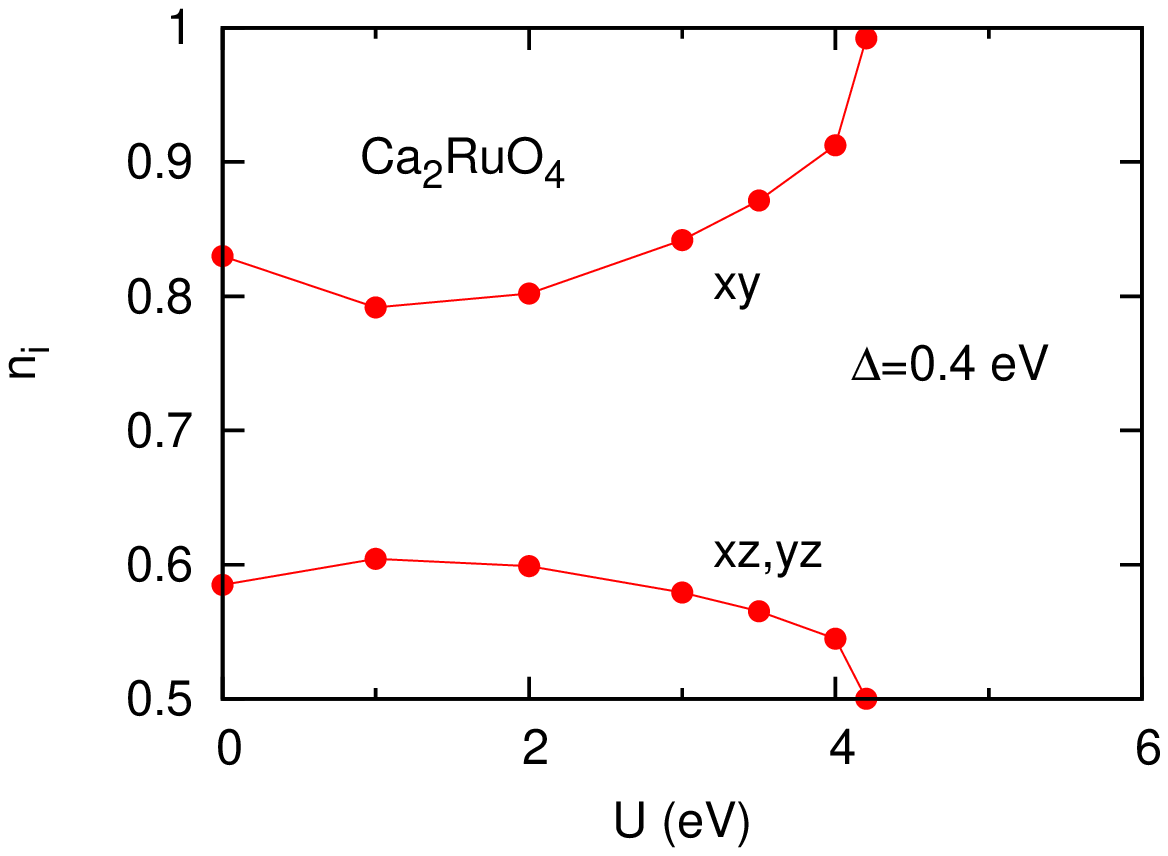}
\end{center}
\caption{(Color online)
Subband occupancies obtained within finite temperature multi-band ED/DMFT 
as functions of $U$.
Upper panels: LaTiO$_3$ ($3d^1$ configuration) for full Hund exchange with 
$J=0.65$~eV\cite{prb08}; 
 V$_2$O$_3$ ($3d^2$ configuration) for full Hund exchange
(blue dots) and Ising-like exchange (red solid dots) with $J=0.7$~eV\cite{v2o3-al}; 
red empty dots: quantum Monte Carlo DMFT results.\cite{keller} 
Lower panel: Ca$_2$RuO$_4$ ($4d^4$ configuration) for full Hund exchange 
with $J=U/4$ and $\Delta=0.4$~eV $t_{2g}$ crystal field splitting.\cite{prl07}
}\end{figure}

During the recent years single-site multi-orbital DMFT has been used 
extensively to investigate the metal insulator transition of a variety 
of materials.\cite{kotliar06} Here, we briefly discuss some of these 
systems which are regarded as typical Mott insulators, and which all 
exhibit characteristic changes of the electronic structure as the metallic 
phase is replaced by the insulator at large Coulomb energies.

Figure~9 shows the correlation driven enhancement of orbital polarization
for LaTiO$_3$, V$_2$O$_3$ and Ca$_2$RuO$_4$.
Because of the orthorhombic structure of LaTiO$_3$, LDA calculations 
reveal that the $a_g$ subbands of the $t_{2g}$ sector are slightly more 
occupied than the two $e'_g$ components.\cite{pavarini} Local Coulomb 
interactions enhance this $t_{2g}$ crystal field splitting,
so that at the Mott transition close to $U=5$~eV the $e'_g$ bands become 
nearly empty and the $a_g$ band half-filled.\cite{pavarini,prb08}   
In the case of V$_2$O$_3$, the corundum lattice structure ensures that 
the doubly degenerate $e'_g$ bands have slightly larger binding energy 
than the $a_g$ bands.\cite{keller} With increasing Coulomb interaction 
this crystal field splitting is strongly enhanced, until in the range 
$U\approx 5\ldots6$~eV, the $a_g$ bands are pushed above the Fermi level 
and the $e'_g$ bands become half-filled.\cite{keller,poteryaev,v2o3-al}  
%
%
Finally, in the case of Ca$_x$Sr$_{2-x}$RuO$_4$, Sr substitution via the 
smaller Ca ions gives rise to an enlarged crystal field splitting between 
$d_{xy}$ and $d_{xz,yz}$ like subbands.\cite{fang} Coulomb correlations 
increase this splitting, until in the Mott phase the $d_{xy}$ bands are 
fully occupied and the half-filled $d_{xz,yz}$ bands are split into lower 
and upper Hubbard bands.\cite{prl07}      

\begin{figure}[t]  
\begin{center}
\includegraphics[width=5.7cm,height=7.5cm,angle=-90]{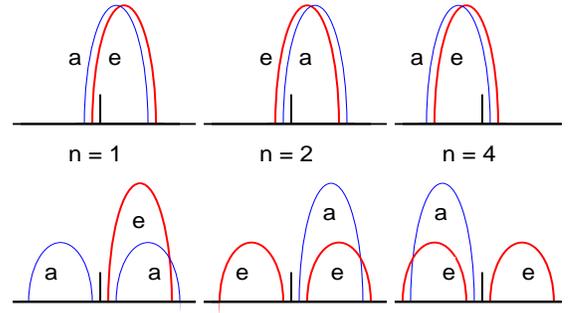}
\end{center}
\vskip-17mm
\caption{(Color online)
Schematic illustration of correlation driven enhancement of orbital 
polarization. 
Upper row: crystal field split $t_{2g}$ LDA densities of states for 
occupancies $n=1$, 2, and 4, corresponding to LaTiO$_3$, V$_2$O$_3$ 
and Ca$_2$RuO$_4$, respectively. Blue curves: singly-degenerate $a_g$ 
band, red curves: doubly-degenerate $e'_g$ bands. In the case of 
Ca$_2$RuO$_4$, $a$ refers to $d_{xy}$, $e$ to $d_{xz,yz}$.
The vertical bars denote the Fermi level.
Lower row: orbitally polarized Mott phase. 
$n=1$: empty $e'_g$ bands,
lower and upper Hubbard peaks of half-filled $a_g$ band;
$n=2$: empty $a_g$ band,
lower and upper Hubbard peaks of half-filled $e'_g$ bands;
$n=4$: filled $a_g$ band,
lower and upper Hubbard peaks of half-filled $e'_g$ bands.
}\end{figure}

Schematically, the uncorrelated densities of states of these transition 
metal oxides and the spectra derived within single-site multi-orbital DMFT for 
realistic Coulomb energies are shown in Figure~10. Note that in the metallic
phase, both orbital symmetries contribute to the spectral weight at the 
Fermi level. In the Mott phase, the gap involes transitions between states 
of opposite symmetry character.

Despite the different subband occupancies of these materials, they exhibit 
a similar correlation driven enhancement of orbital polarization.\cite{others}
By pushing some subbands above or below the Fermi level, the effective degeneracy 
is reduced from three to two or one. The Mott transition therefore occurs at 
lower critical Coulomb energy than in a cubic environment with equivalent subbands. 

According to Figure~1, the cluster molecular orbital densities of the 
single band Hubbard model in a multi-site picture also exhibit a 
substantial splitting relative to the total band width. Nevertheless,
Coulomb correlations in these cases only lead to moderate charge transfer
between these cluster orbitals, as demonstrated by the results given in
Fig.~2. The main physical reason for this qualitative difference with
respect to the multi-orbital materials is the strong single-particle 
hybridization among orbitals so that any tendency towards orbital selective 
Mott transitions is suppressed. For the same reason, partial band filling 
or emptying, with a Mott transition in the remaining subset of bands, as 
found in several multi-orbital materials, is also absent.         

A certain amount of inter-orbital hybridization exists also in multi-orbital
systems since the single-electron Hamiltonian in the orbital basis is not 
diagonal throughout the Brillouin Zone. This residual coupling, however,  
is much weaker than in the single-band multi-site system, so that Coulomb 
correlations can indeed lead to nearly complete orbital polarization. This is 
supported by the DMFT results for LaTiO$_3$ and V$_2$O$_3$ which have been
studied both by evaluating the lattice Green's function via Eq.~(\ref{G}) (see
Refs.\cite{pavarini,poteryaev}) and via the approximate version, Eq.~(\ref{Grho})
(see Refs.\cite{prb08,keller,v2o3-al}). In these systems both formulations give 
very similar results, in particular, both confirm the scenario of strong orbital
polarization. 

\bigskip

\section{Conclusion}

Cellular DMFT combined with 
finite temperature exact diagonalization has been used to investigate the
influence of short range correlations on the Mott transition in the single-band
Hubbard model. Both square and triangular lattices at half-filling were studied.
A mixed basis consisting of cluster sites and bath molecular orbitals was shown
to provide an efficient method for the evaluation of the cluster self-energies
and Green's functions. Since in the cluster molecular orbital representation
these quantities become diagonal, an intriguing analogy exists between Coulomb
correlations in these multi-site single-band systems and several multi-orbital
materials which were studied previously within single-site DMFT. 

In remarkable
contrast to LaTiO$_3$, V$_2$O$_3$, and Ca$_2$RuO$_4$, which exhibit pronounced
orbital polarization at the Mott transition, the single-band systems
show very little correlation driven enhancement of orbital polarization.
Thus, all cluster molecular orbitals take part in the metal insulator transition.
Moreover, the transition occurs at the same critical $U$ for all cluster orbitals. 
Since an approximate momentum variation of the lattice self-energy and Green's
function can be constructed from a superposition of these molecular orbital
components, this finding yields the important result that the Mott gap opens
simultaneously across the entire Fermi surface. Thus, for both square and 
triangular lattices at half filling, there is no orbital selective Mott 
transition, where certain sections of the Brillouin Zone would open a gap
at lower Coulomb energy than other parts. Moreover, there is no evidence for 
the combination of subband-filling and Mott transition in other subbands,
that is characteristic of the multi-orbital materials mentioned above. 
It would be of great interest to investigate whether these findings also 
hold at a finer momentum resolution which would require cluster sizes larger 
than $n_c=4$.  

\bigskip
               
{\bf Acknowledgements}\ \ 
We like to thank A. Georges, N. Kawakami, G. Kotliar, R.H. McKenzie, M. Potthoff, 
B.J. Powell and H. Tsunetsugu for useful discussions. The computational work was
carried out on the J\"ulich JUMP computer.


\begin{thebibliography}{99}
\bibitem{others}
   Note, however, other trends such as those discussed in Refs.
   \cite{lechermann,ray,tetra}.
\end{thebibliography}
\end{document}